\documentclass[epj-spec]{svjour}
\usepackage{graphicx}
\usepackage{ulem}
\usepackage{cite,./mcite}
\begin{document}
\title{Standard Model backgrounds to supersymmetry searches}
\author{Michelangelo
  L. Mangano\inst{1}\fnmsep\thanks{\email{michelangelo.mangano@cern.ch}} }
\institute{CERN, PH Department, TH Group, 1211 Geneva 23,
  Switzerland }
\abstract{
This work presents a review of the Standard Model sources of
backgrounds to the search of supersymmetry signals. Depending on the
specific model, typical signals may include jets, leptons, and
 missing transverse energy due to the escaping lightest supersymmetric
 particle. We focus on the simplest case of multijets and missing
 energy, since this allows us to expose most of the issues common to
 other more complex cases. The review is not exhaustive, and is aimed
 at collecting a series of general comments and observations, to serve
 as guideline for the process that will lead to a complete
 experimental determination of size and features of such SM processes.
} 
\maketitle

\def    \be             {\begin{equation}}
\def    \ee             {\end{equation}}
\def    \ba             {\begin{eqnarray}}
\def    \ea             {\end{eqnarray}}
\def    \nn             {\nonumber}
\def    \=              {\;=\;}
\def    \frac           #1#2{{#1 \over #2}}
\def    \ret            {\\[\eqskip]}
\def    \ie             {{\em i.e.\/} }
\def    \eg             {{\em e.g.\/} }
\def \lsim{\mathrel{\vcenter
     {\hbox{$<$}\nointerlineskip\hbox{$\sim$}}}}
\def \gsim{\mathrel{\vcenter
     {\hbox{$>$}\nointerlineskip\hbox{$\sim$}}}}
\def    \ipb            {\mbox{$\mathrm{pb}^{-1}$}}
\def    \ev            {\mbox{$\mathrm{eV}$}}
\def    \kev            {\mbox{$\mathrm{keV}$}}
\def    \mev            {\mbox{$\mathrm{MeV}$}}
\def    \gev            {\mbox{$\mathrm{GeV}$}}
\def	\tev		{\mbox{$\mathrm{TeV}$}}
\def    \pt             {\mbox{$p_{\mathrm T}$}}
\def    \et             {\mbox{$E_{\mathrm T}$}}
\def    \etj             {\mbox{$E_{\mathrm T,j}$}}
\def    \met            {\mbox{$\rlap{\kern.2em/}E_T$}}
\def    \meff            {\mbox{$M_{\mathrm{eff}}$}}
\def    \ifb            {\mbox{$\mathrm{fb}^{-1}$}}
\def    \as             {\ifmmode \alpha_s \else $\alpha_s$ \fi}
\def    \asq             {\ifmmode \alpha_s(Q) \else $\alpha_s(Q)$ \fi}

\section{Introduction}
The Standard Model (SM) of fundamental interactions has by now been
successfully tested over the past 30 years, validating its dynamics
both in the gauge sector, and in the flavour structure, including a
compelling confirmation of the source of the observed violation of
parity (P) and combined charge and parity (CP) symmetries. The
inability of the SM to account for established features of our
universe, such as the presence of dark matter, the baryon asymmetry,
and neutrino masses, are not considered as flaws of the SM, but as
limitations of it, to be overcome by adding new elements, such as new
interactions and new fundamental particles.  With this perspective,
the LHC is not expected to further test the SM, but to probe, and
hopefully provide evidence for, the existence of such new
phenomena. Our ability to predict what will be observed at the LHC is
therefore not limited by fundamental issues related to left-over
uncertainties about the SM dynamics, but by the difficulty of
mastering the complex strong-interaction dynamics that underlies the
description of the final states in proton-proton
collisions~\cite{Campbell:2006wx}.

Many years of experience at the Tevatron collider, at HERA, and at
LEP, have led to an immense improvement of our understanding of this
dynamics, and put us today in a solid position to reliably anticipate
in quantitative terms the features of LHC final states. LEP, in
addition to testing with great accuracy the electroweak interaction
sector, has verified at the percent level the predictions of
perturbative QCD, from the running of the strong coupling constant, to
the description of the perturbative evolution of single quarks and
gluons, down to the non-perturbative boundary where strong
interactions take over and cause the confinement of partons into
hadrons. The description of this transition, relying on the
factorization theorem that allows to consistently separate the
perturbative and non-perturbative phases, has been validated by the
comparison with LEP data, allowing the phenomenological parameters
introduced to model hadronization to be determined. The factorization
theorem supports the use of these parameters for the description of
the hadronization transition in other experimental environments. HERA
has made it possible to probe with great accuracy the short-distance
properties of the proton, with the measurement of its partonic content
over a broad range of momentum fractions $x$. These inputs, from LEP
and from HERA, beautifully merge into the tools that have been
developed to describe proton-antiproton collisions at the Tevatron,
where the agreement between theoretical predictions and data confirms
that the key assumptions of the overall approach are robust. Basic
quantities such as the production cross section of $W$ and $Z$ bosons,
of jets up to the highest energies, and of top quarks, are predicted
theoretically with an accuracy consistent with the known experimental
and theoretical systematic uncertainties. This agreement was often
reached after several iterations, in which both the data and the
theory required improvements and reconsideration.  See, for example,
the long saga of the bottom-quark cross section~\cite{Mangano:2004xr}\,,
or the almost embarrassing --- for theorists --- case of the production
of high transverse momentum $J/\psi$s~\cite{Abe:1992ww}\,.  

While the
present status encourages us to feel confident about our ability to
extrapolate to the LHC, the sometimes tortuous path that led to this
success demands caution in assuming by default that we know all that
is needed to accurately predict the properties of LHC final
states. Furthermore, the huge event rates that will be possible at the
LHC, offering greater sensitivity to small deviations, put stronger
demands on the precision of the theoretical tools. In this essay I
discuss the implications of these considerations, using as a specific
example the search for supersymmetry. I
shall not provide a systematic discussion of all search strategies,
which vary a lot in their details depending on the specific version of
the assumed supersymmetry breaking patterns, and of the parameters
values. I shall rather focus on the canonical case of the multijet
plus missing transverse energy signals, which are well suited to
address the rather general background-estimate issues which are the
focus of this note. Likewise, I shall
not review the state of the art in
calculations and Monte Carlo tools (for these, see
Refs.~\cite{Dixon:2007hh} and~\cite{Dobbs:2004qw}, respectively) but
will confine myself to the applications of several tools that have
recently been developed, and their impact on the expected signals.

For more general considerations on the issue of ``discovering new
physics at the LHC'', I refer the reader to the recent essay in
ref.~\cite{Mangano:2008ag}. 

\section{Signal properties}
To be concrete, I will consider the production and decay of strongly
interacting supersymmetric particles, like gluinos and squarks. Chain
decays of such particles lead to final states such as those shown in
fig.~\ref{fig:gluinodec}. Pair production will therefore typically
lead to configuration with several jets, missing transverse energy,
and possibly leptons~\cite{Barnett:1987kn}. 
I shall focus on the case of
no-leptons~\cite{Baer:1995nq}. 
Final states with leptons~\cite{Baer:1995va,Baer:2008kc} 
have backgrounds similar
to those studied here, with the addition of possible extra gauge
bosons, as well as contributions
coming from charm and bottom quarks, which occasionally lead to
isolated leptons~\cite{Barbieri:1991vk,Sullivan:2008ki}. 
 There is a vast literature covering all these facets, and I
just refer to the experimental literature for an overview of the state
of the art in the current searches at the
Tevatron~\cite{:2007ww,Affolder:2001tc},  and of the
prospects for discovery at the LHC~\cite{Hinchliffe:1996iu,AtlasTDR,
  Abdullin:1998pm, Ball:2007zza, Allanach:2002nj,
  AguilarSaavedra:2005pw, Spiropulu:2008ug}.

\begin{figure}
\begin{center}
\includegraphics[width=0.99\textwidth]{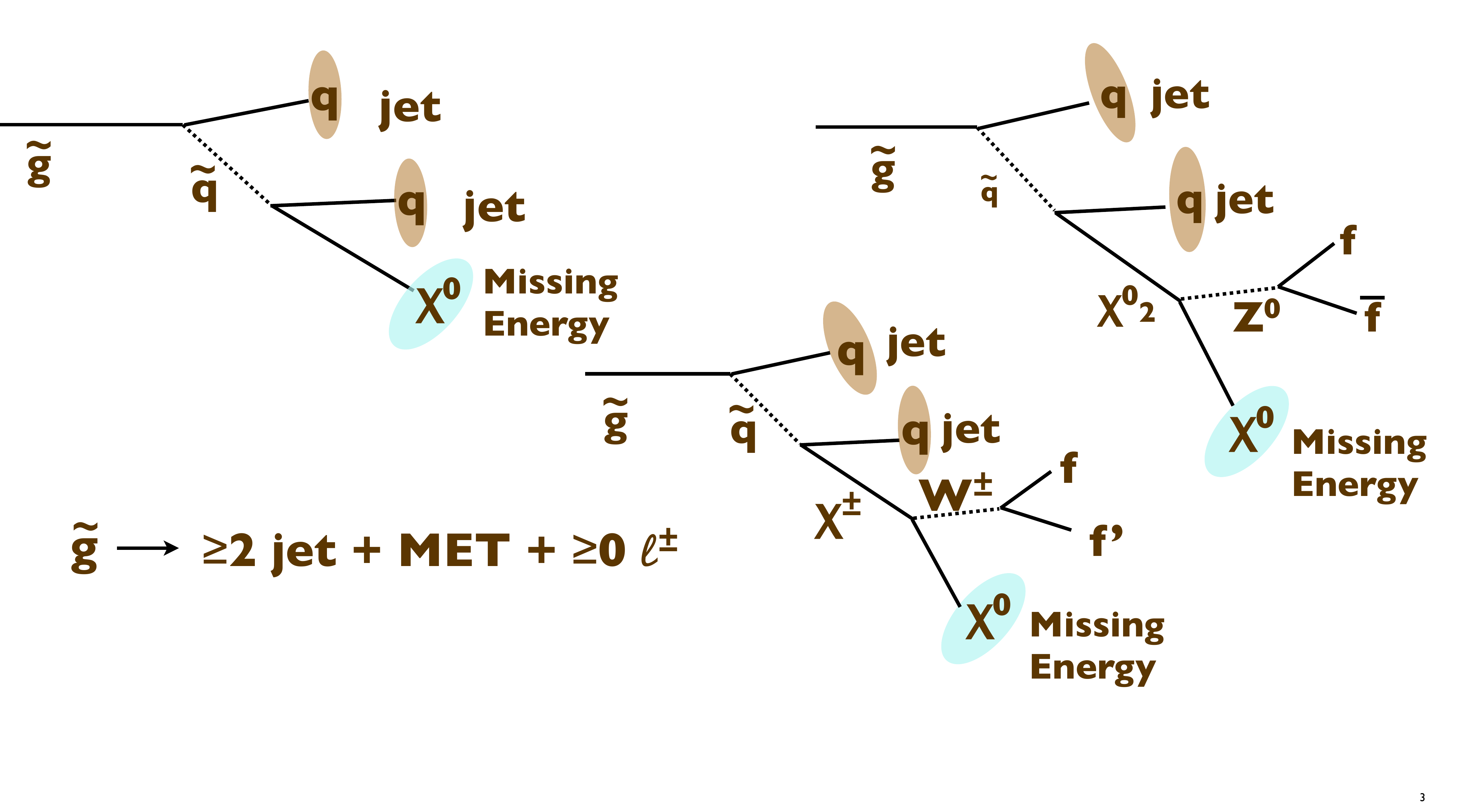}
\caption{Possible topologies for the final states of gluino decays.}
\label{fig:gluinodec}
\end{center}
\end{figure}

The cuts used in the following are the 
typical analysis cuts chosen by the ATLAS
experiment~\cite{Vahsen,Yamazaki:2008nm,Yamamoto:2007it}  
to optimize the signal extraction:
\ba
&&
N_{jets}\ge 4, \quad \mbox{with} \quad \et>50~\gev~~\mbox{for all
  jets, and}~~\et>100~\gev~~\mbox{for the leading jet;} \nn \\
&&
\mbox{no lepton with}~~\et>20~\gev \nn \\
&&
\met>\max(100~\gev,0.2\times\meff) \\
&&
\label{eq:susycuts}
\meff=\met+\sum_{\mathrm{jets}} \etj \nn \\
&&
\mbox{Transverse sphericity,}~~S_T > 0.2 \nn
\ea
A sample result, 
obtained from the full detector simulation of the
backgrounds and of the signal, for a specific point in supersymmetric
parameter space, is shown in fig.~\ref{fig:susy}.
\begin{figure}
\begin{center}
\includegraphics[width=0.9\textwidth]{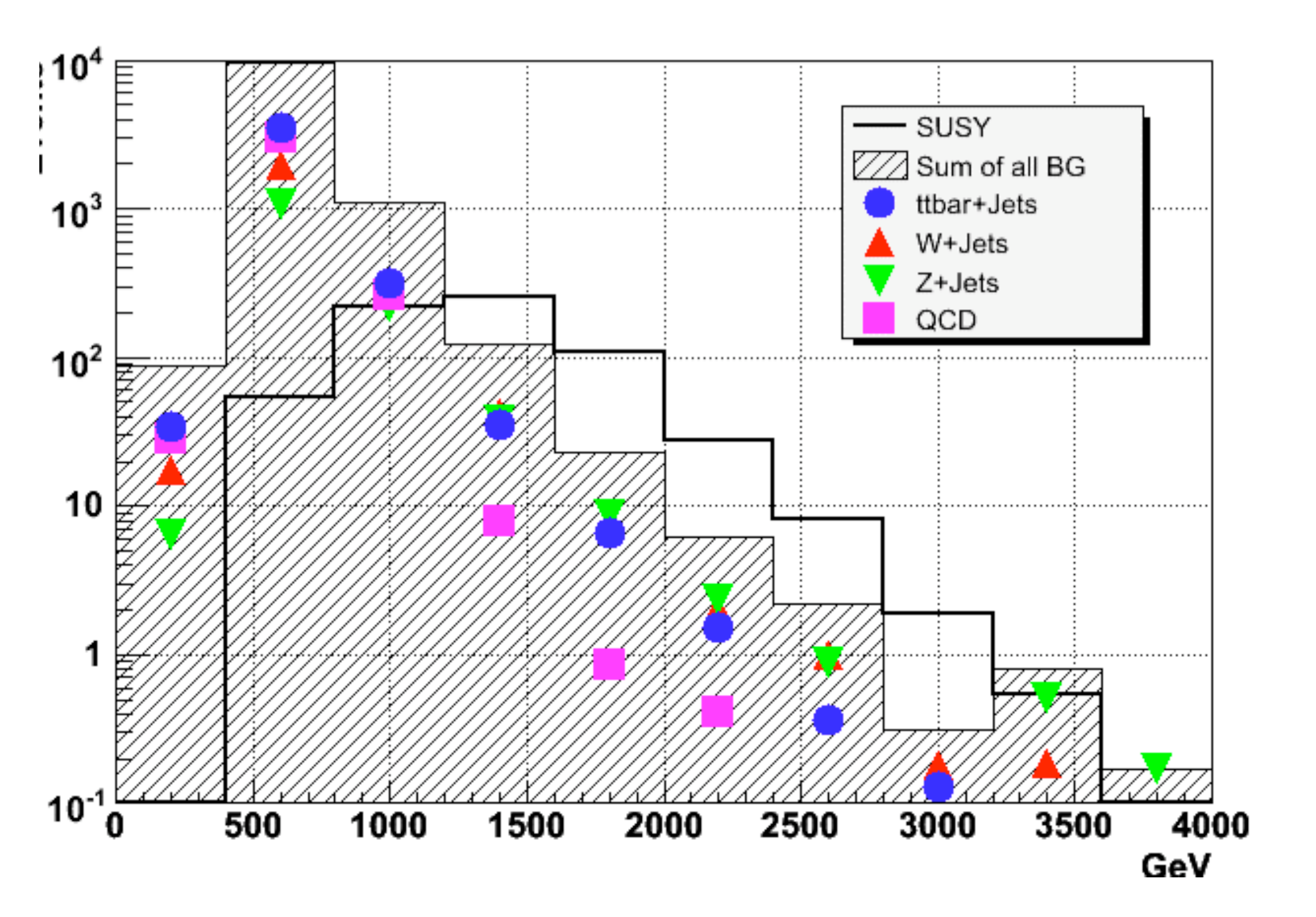}
\caption{Example of an expected supersymmetry signal and relative 
backgrounds in the
 multijet+missing transverse energy final state~\protect\cite{Vahsen}\,.}
\label{fig:susy}
\end{center}
\end{figure}
The signal corresponds to production of squarks and gluinos with a
mass of the order of 1~TeV, in a typical so-called mSUGRA model,
defined by the values of few parameters at the scale of grand
unification. The renormalization-group evolution to the electroweak
scale leads to the breaking of the gauge group $SU(2)\times U(1)$, and
defines the entire spectrum of the supersymmetric partners. The
parameters are given by the common mass $m_0$ (400~GeV in our example)
of the scalar partners
of the Standard Model fermions, by the common mass $m_{1/2}$ (400~GeV) of the
fermionic partners of the gauge bosons, by the mixing $\tan\beta$ (10)
between the expectation values of the two Higgs doublets, by the sign
of the mass
term $\mu$ of the Higgs fields ($\mu>0$), 
and by the parameter $A$ (0) defining the
soft-supersymmetry-breaking scalar potential. The resulting gluino
mass is $\sim$950~GeV, and the mass of the scalar partners of the
light quarks is $\sim$930~GeV.

The results of fig.~\ref{fig:susy} are rather general in this class of models:
varying the parameters that determine the gluino and squark masses
will shift the signal to lower (higher) values of the effective mass
 \meff, and to higher (lower)
rates, depending on whether the resulting sparticle masses are reduced
or increased. While the quantitative details of the following analysis
will depend on the specific set of chosen parameters, the spirit of
the analysis remains the same. 

Notice that 
while the signal has certainly a statistical significance sufficient
to claim a deviation from the SM, it is unsettling that its shape is
so similar to that of the sum of the backgrounds. The theoretical
estimates of these backgrounds have also increased significantly over the
last few years, as a result of more accurate tools to describe
multijet final states. There is no question, therefore, that unless
each of the background components can be separately tested and
validated, it will not be possible to draw conclusions from the mere
comparison of data against the theory predictions. 

I am not saying this because I do not believe in the goodness of our
predictions. But because claiming that supersymmetry exists is far too
important a conclusion to make it follow from the straight comparison
against a Monte Carlo. One should not forget relevant examples from
the colliders' history~\cite{Arnison:1984iw,Arnison:1984qu}\,, such as
the misinterpretation in terms of top or supersymmetry of final states
recorded by UA1 with jets, \met, and, in the case of top,
leptons. Such complex final states were new experimental
manifestations of higher-order QCD processes, a field of phenomenology
that was just starting being explored quantitatively. It goes to the
theorists' credit to have at the time played devil's
advocate~\cite{Ellis:1985ig}\,, and to have improved the SM predictions,
to the point of proving that those signals were nothing but bread and
butter $W$ or $Z$ plus multijet production. But the fact remains that
claiming discoveries on the basis of a comparison against a MC is dangerous.

The lesson for the future is that, more than accurate theoretical
calculations, in these cases one primarily needs a strategy for an internal
validation of the background estimate. If evidence for some new phenomenon
entirely depends on the shape of some distribution, however accurate
we think our theoretical inputs are, the conclusion that there is new
physics is so important that people will always correctly argue
that perhaps there is something weird going on on the theory side, and
more compelling evidence has to be given. 

In what follows I review the nature of the backgrounds, the status of
their theoretical understanding, and the possible approaches to
determine them directly from the data.

\section{ Background classification}
It is useful to classify backgrounds in three categories: irreducible,
reducible, and instrumental. 

{\it Irreducible} backgrounds are those that, on an event by event
basis, cannot be distinguished from the signal, even in presence of a
perfect detector. In the case of the supersymmetric signals we are
discussing, they emerge from production of $Z$ plus 4 jets, with the
$Z$ boson decaying to invisible neutrinos. While the missing mass of
the event should reconstruct $M_Z$, the tail of the Breit-Wigner
distribution can generate events beyond the kinematical range allowed
by the neutralino masses. This problem is enhanced by the lack of
information on the longitudinal component of the missing-momentum
vector.

{\it Reducible} backgrounds include processes that share the main
features of the signal, but have in addition some extra element that
would make them in principle distinguishable from it. The exploitation
of these additional elements may be limited, however, either because
of the need to maintain a good signal efficiency, or due to a limited
experimental sensitivity to the distinguishing elements. Examples of
relevance to our case study include:
\begin{itemize}
\item $W+3$~jets, with $W\to \tau\nu$ and the hadronic decay of the
  $\tau$ giving rise to the fourth hadronic jet required in the event
  selection. The $\tau$ can in principle be identified, and rejected,
  by using both the low multiplicity of the resulting jet, and
  because, as a result of the $\tau$ lifetime, 
  the charged tracks of the jet originate from a vertex displayed
  relative to the primary vertex. The efficiency for tagging such
  $\tau$ decays is however limited, and a residual background will be
  left. Notice that, due to the presence of the neutrinos from both
  the $W$ and the $\tau$ decays, in addition to the lack of
  information on the longitudinal momentum of the neutrinos, the
  missing mass is also poorly determined, and therefore cannot be used
  on an event-by-event basis to suppress the background.
\item $W+4$~jets, with the $W$ decaying to an undetected lepton and
  a neutrino. The efficiency to detect, and reject, the leptons,
  cannot be perfect, since leptons can have low-\pt. This is
  particularly true of $W$ decays to $\tau$, with the $\tau$ itself
  decaying leptonically.
\item $t\bar{t}$ events contain $W$ bosons and jets, and therefore
  fall automatically under the above categories. The additional
  presence of $b$ quarks provides a possible handle to reject them,
  but vetoing on $b$ quarks could be an unwise choice, since the
  signal itself could have an enhanced fraction of $b$, for example if
  third-generation squarks were lighter than those of the first two
  generations. The large rate of $t\bar{t}$ production at the LHC
  makes this process a potentially large reducible background.
\end{itemize}

{\it Instrumental} backgrounds arise when the characteristic features
of the signals are due to the inaccuracy of the detector or of the
measurement. The most important example in this category is QCD
multijets, namely final states with only jets. The missing transverse
momentum is entirely a result of either of the following three
effects: 
\begin{itemize}
\item A mismeasurement of the energy of the individual jets, leading
  to an overall imbalance of transverse energy
\item The incomplete coverage of the calorimeters, which could allow
  some jets to escape reconstruction, thus leading once again to a
  momentum imbalance.
\item Accidental extra deposits of energy, like cosmic rays hitting
  part of the detector in time with the recording of the event,
  backgrounds from protons in the halo of the beam colliding with some
  detector element upstream of the interaction region, large
  fluctuations of electronic
  noise in the calorimeters, etc.
\end{itemize}
The size of the multijet cross section is huge, much bigger than any
possible signal, and therefore even a small contamination of the
missing energy measurements can lead to significant backgrounds. As a
reference, we show in fig.~\ref{fig:qcdnomet}
 the \meff\ distribution of 4-jet final states from QCD, with jets passing the
 selection cuts of eq.~\ref{eq:susycuts}, but without any \met\
 requirement. 
\begin{figure}
\begin{center}
\includegraphics[width=0.9\textwidth]{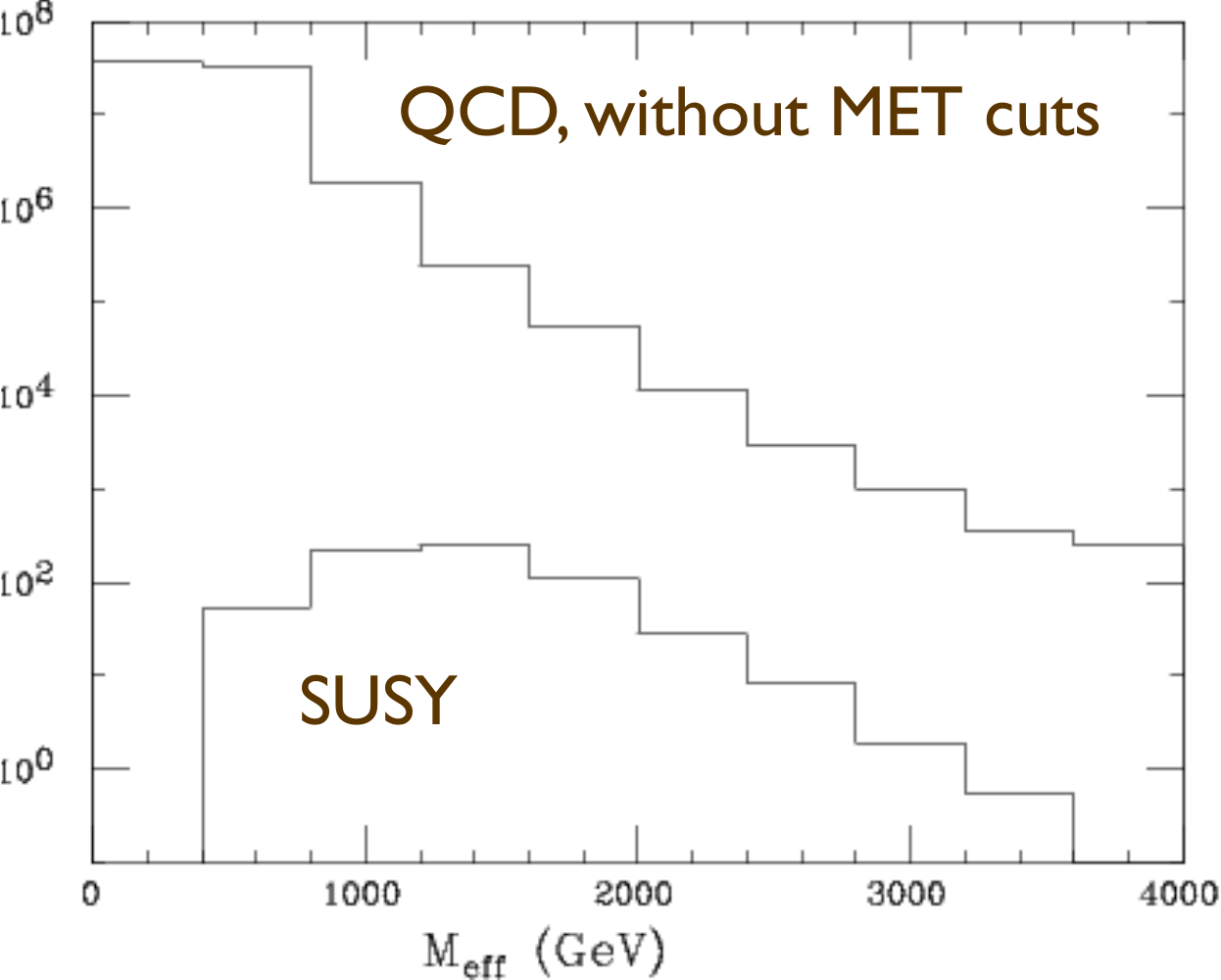}
\caption{QCD multijet rates as a function of the effective mass
  variable \meff, defined in eq.~(\ref{eq:susycuts}), in events
  without missing transverse energy requirement.}
\label{fig:qcdnomet}
\end{center}
\end{figure}
To highlight the difficulty of properly removing unwanted sources of
\met, we show in fig.~\ref{fig:cdfmet} the raw spectrum of jet \et\
and of \met, at D0 and CDF, 
before and after the removal of events with an identified spurious
source of energy deposits. The lower rate of cosmic rays penetrating
to the depth of the LHC experiments, the shorter time window allowed
by the much higher repetition rate of LHC collisions, and the higher
accuracy of the calorimeters, will certainly significantly reduce this
problem at the LHC, but the size of the effects certainly poses very
difficult experimental challenges. 
 
\begin{figure}
\begin{center}
\includegraphics[width=0.55\textwidth]{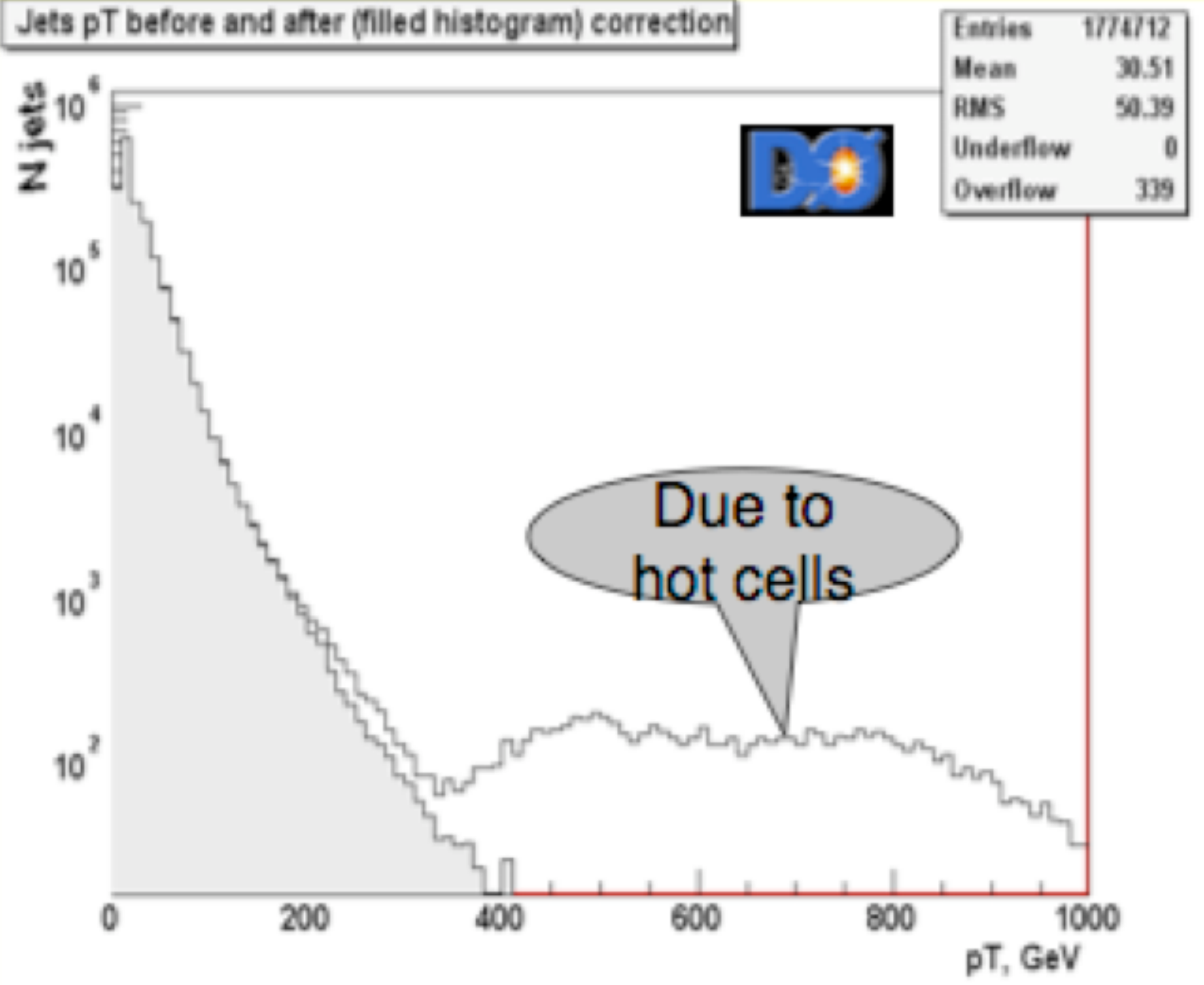} \hfill
\includegraphics[width=0.4\textwidth]{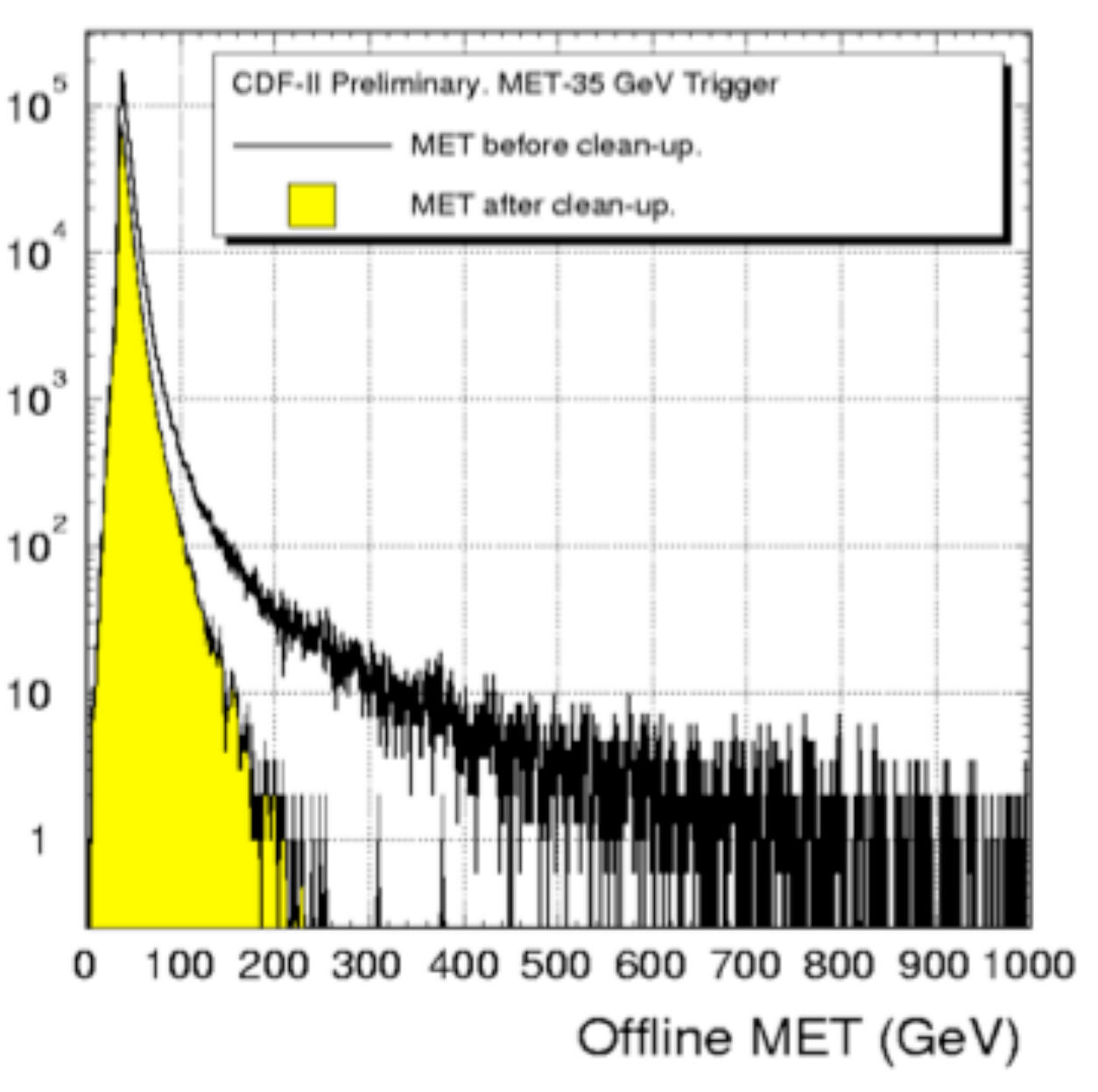}
\caption{Left: the jet \et\ spectrum in D0, before and after removal
  of the so-called {it hot-cells}, namely additional detector noise
  adding energy to the jet reconstruction. Right: the missing \et\
  spectrum at CDF, before and after the clean-up of all sources of
  spurious energy deposits.}
\label{fig:cdfmet}
\end{center}
\end{figure}

The prediction for each of these backgrounds, whether their origin is
physics or detector effects, as well as possible additional ones, can
only partly rely on our a-priori simulation of the physical processes
and of the detector performance. The conclusion that a signal for new
physics has been identified, to be credible, will have to mostly rely
on the data themselves. Each search strategy should therefore contain
the definition of control samples and control observables to be used
for the direct determination of the backgrounds, e.g. by extrapolating
the sidebands of a given distribution, or by validating the MC
simulation tools that will be used to extrapolate our knowledge from
the control sample to the signal region. Furthermore,  efforts should be
made to establish that the extrapolation of the knowledge acquired
from the control samples to the signal region is legitimate and
reliable.
For interesting reviews of how the experiments plan to utilize their
data in the process of assessing the size of the backgrounds, see
e.g. ref.~\cite{Yamazaki:2008nm,Yamamoto:2007it}.

\subsection{Theoretical tools for SM backgrounds}
A rich arsenal of  theoretical tools is available today to address
these tasks. Next-to-leading-order (NLO) calculations exist for
processes of interest with up to 3 final state particles, and are
available in the form of codes that allow to implement basic analysis
cuts at the partonic level, such as jet \et\ thresholds, or rapidity
cuts: production
of 2~\cite{Ellis:1985er,Ellis:1990ek, Aversa:1990uv,
  Ellis:1994dg,Giele:1994gf,Frixione:1997np} 
and 3 jets~\cite{Frixione:1995ms,Kilgore:1996sq,Nagy:2003tz}, heavy quark
pairs~\cite{Nason:1987xz,Beenakker:1988bq,Mangano:1991jk} 
(possibly with an extra jet~\cite{Dittmaier:2008jg}),
associated production of electroweak gauge bosons and one or two
jets~\cite{Campbell:2002tg,Campbell:2003hd}, possibly including heavy
quarks~\cite{Campbell:2006cu,Campbell:2005zv}.
Next-to-next-to-leading-order (NNLO) results are available for lower
multiplicities final states, most notably inclusive
 Drell-Yan~\cite{Hamberg:1990np} and Higgs 
production~\cite{Harlander:2002wh,Anastasiou:2002yz}. 
Also in this case, calculations exist allowing for explicit cuts to be
placed on the final state particles, simulating more closely the
impact of experimental 
analyses~\cite{Anastasiou:2003ds,Melnikov:2006di,Anastasiou:2007mz,
Grazzini:2008tf}.

Such parton-level calculations make it possible to predict inclusive
quantities, and to accurately benchmark the absolute production
rates. This serves the dual purpose of assessing the stability and
reliability of the leading-order (LO) 
calculations typically used in the full event
generators, and of allowing the best possible determination of the
properties (cross-sections and couplings) of the new particles being
discovered. Several developments~\cite{Frixione:2002ik,Nason:2004rx,
  Frixione:2007vw} have also solved the challenge of
incorporating exact NLO calculations in the shower Monte Carlos~\cite{
  Corcella:2000bw, Sjostrand:2003wg}, leading to complete NLO shower
MC codes~\cite{Frixione:2006gn,Frixione:2003ei,Frixione:2007nw}. 

The backgrounds to processes such as supersymmetric particle
production, on the other hand, feature the presence of many jets in
the final states. NLO calculations for such final states are still
beyond feasibility, and one needs to resort to
event generators capable of describing final
states with large jet multiplicity, and of returning events where the
full set of final hadrons, including the description of the evolution
of the proton fragments, in order to simulate as accurately as
possible the way a given class of processes will appear inside the
detector. The goal in this case is not necessarily a first-principle
predictivity, but to achieve, possibly after tuning against the data, a
good agreement with the data. 

To achieve this, our calculations need to describe as accurately as
possible both the full matrix elements for the underlying hard
processes, as well as the subsequent development of the hard partons
into jets of hadrons.  However, for the complex final-state topologies
we are interested in, no factorization theorem exists to rigorously
separate these two components.  The main obstacle is the existence of 
several hard scales, like the jet transverse energies and di-jet 
invariant masses, which for a generic multi-jet event will span
a wide range.  This makes it difficult to unambiguously separate the
components of the event, which belong to the ``hard process'' (to be
calculated using a multi-parton amplitude) from those developing
during its evolution (described by the parton shower).  A given
$(n+1)$-jet event can be obtained in two ways: from the
collinear/soft-radiation evolution of an appropriate $(n+1)$-parton
final state, or from an $n$-parton configuration where hard,
large-angle emission during its evolution leads to the extra jet.  A
factorization prescription (in this context this is often called a
``matching scheme'' or ``merging
scheme'') 
defines, on an
event-by-event basis, which of the two paths should be followed.  The
primary goal of a merging scheme is therefore to avoid double counting
(by preventing some events to appear twice, once for each path), as
well as dead regions (by ensuring that each configuration is generated
by at least one of the allowed paths).  Furthermore, a good merging
scheme will optimize the choice of the path, using the one that
guarantees the best possible approximation to a given kinematics. 
Different merging schemes have been
proposed~\cite{Catani:2001cc,Lonnblad:2001iq,Krauss:2002up, 
  Mangano:2006rw, Mrenna:2003if ,Alwall:2007fs}, all
avoiding the double counting and dead regions, but leading to
different results in view of the different ways the calculation is
distributed between the matrix element and the shower evolution.  As in
any factorization scheme, the physics is independent of the separation
between phases only if we have complete control over the perturbative
expansion. Otherwise a residual scheme-dependence is left. Exploring
different merging schemes is therefore one of the elements necessary 
to assess the
systematic uncertainties of multi-jet calculations.

In the next sections we shall review, one by one, the issues that
emerge when considering each of the background sources introduced
above, and the current status of the theoretical predictions,
describing, where possible, the validation tests that can be
performed today either comparing different calculations, or comparing
against the available Tevatron data.

\section{$Z$+jets}
We start from the only irreducible background, namely
$(Z\to\nu\bar{\nu})+$4~jets. Its size, compared to the supersymmetry
signal extracted from fig.~\ref{fig:susy}, is shown in
fig.~\ref{fig:z4j}, after the sequential application of the analysis
cuts. This comparison, as well as those that will be shown in the
subsequent sections, is only indicative since, while the signal is
derived from a full detector simulation, the background was obtained
by applying simple particle-level cuts to the output of the event
generator~\cite{Mangano:2002ea}. 
The main purpose of this study is therefore to show the
relative size of the different background components, and their
evolution as a function of the cuts.

The uppermost histogram corresponds to the application of just the
\et\ cuts to the 4 jets. The following curves correspond to imposing the
cut to the \met (namely to the \pt\ of the $Z$ boson), and to the
transverse sphericity, $S_T$. No veto against leptons is required,
since there are no leptons in this final state.
\begin{figure}
\begin{center}
\includegraphics[width=0.99\textwidth]{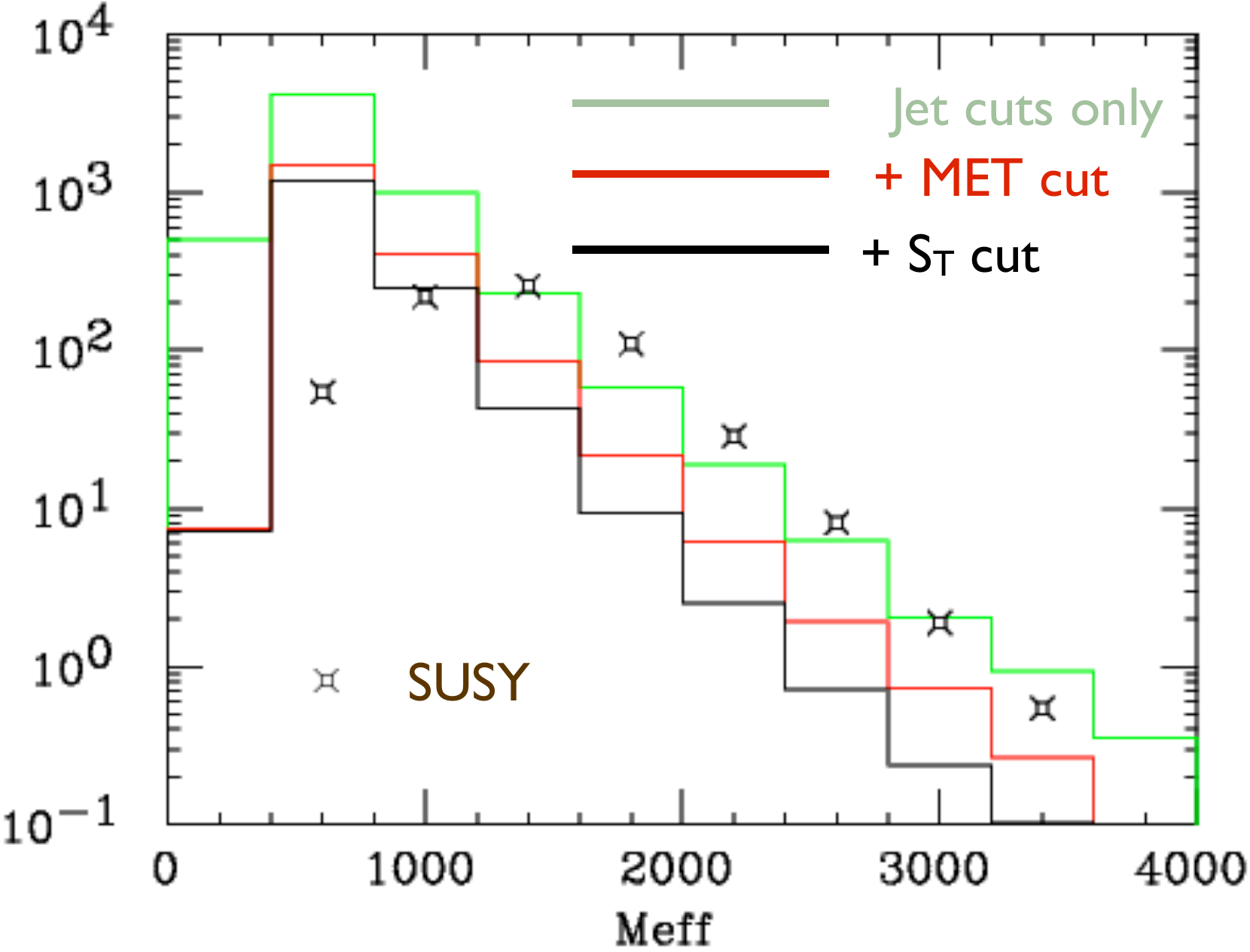}
\caption{Distribution of the \meff\ variable for the $Z$+4~jet
  process, with $Z\to\nu\bar{\nu}$. The different histograms represent
  the evolution of the background when additional signal cuts are
  imposed to the final state.}
\label{fig:z4j}
\end{center}
\end{figure}
Notice that, while the absolute background rate after all cuts is only
about 10\% of the signal, the shapes of the two in the large-\meff\
region are very similar. How reliable is this absolute normalization?
Figure~\ref{fig:z4jmc} shows the important difference in rate obtained
using a shower MC to generate the jets from the shower evolution, and
using a full matrix-element calculation. As the jet multiplicity grows,
the shower approach underestimates the cross section more and
more. While the matrix-element approach is certainly more reliable,
this comparison underscores the possibly large systematics of a
theoretical calculation for such complex final states. The validation
against the data is therefore a necessary step before any application
of the theoretical predictions. 
\begin{figure}
\begin{center}
\includegraphics[width=0.99\textwidth]{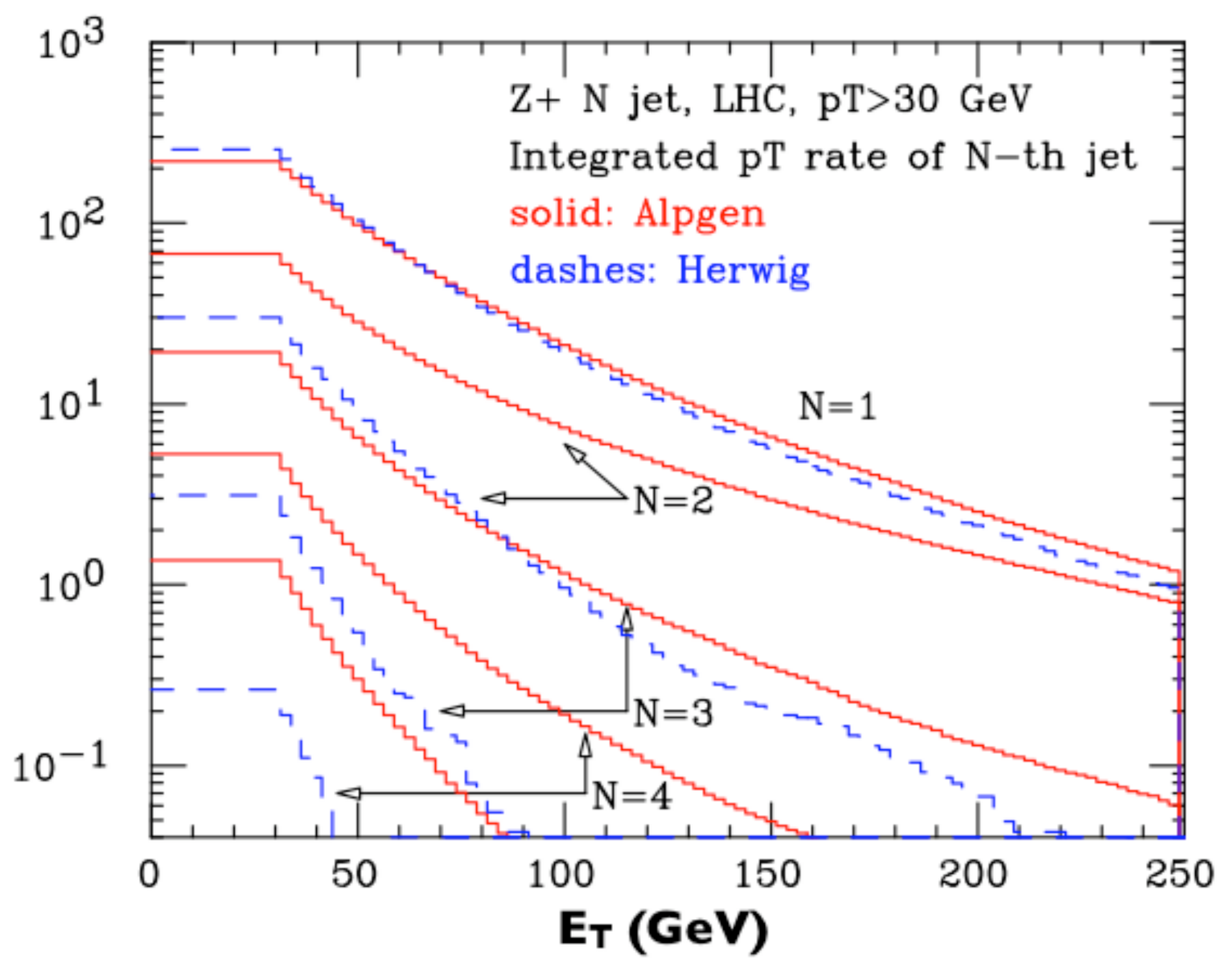}
\caption{Integrated \et\ spectrum of the $N$-th jet ($N=1,2,3,4$) in
  $Z+N$-jet events, as derived from the shower monte carlo HERWIG (dashes) and
  from a leading-order matrix-element calculation with the monte carlo
 ALPGEN (solid).}
\label{fig:z4jmc}
\end{center}
\end{figure}
Confidence in the reliability of the matrix element calculations comes
from the measurements of the Tevatron
experiments of $Z$+multijet final states, with $Z\to \ell^+\ell^-$. 
CDF~\cite{Aaltonen:2007cp}
 has compared
the $Z+1$ and 2 jet rates against the NLO calculations
of~\cite{Campbell:2002tg}, finding excellent agreement in both
normalization and in shapes, as shown in fig.~\ref{fig:CDFZjet}. The cross
section for multijet production has been measured by
D0~\cite{Abazov:2006gs}, and compared against the NLO results for 1
and 2 jets, and against LO matrix
elements~\cite{Mrenna:2003if} 
for up to 4 jets, as shown
in fig.~\ref{fig:D0Zjet}

\begin{figure}
\begin{center}
\includegraphics[width=0.99\textwidth]{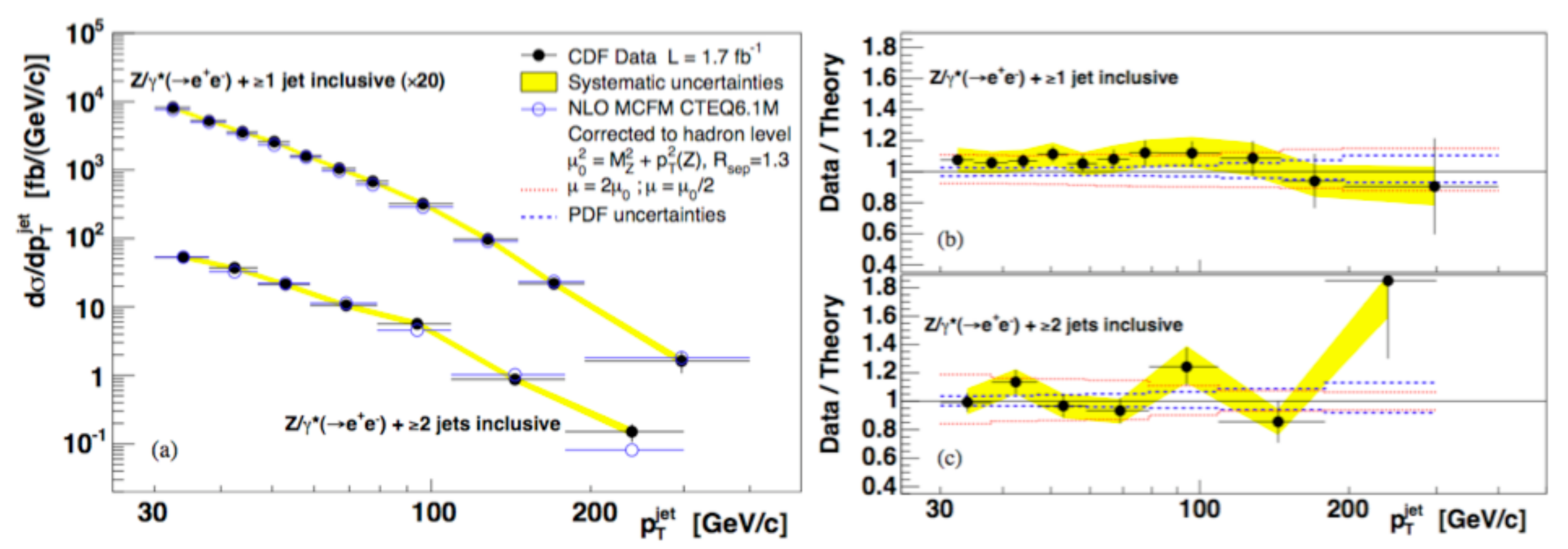}
\caption{Leading-jet \et\ spectrum in $Z+\ge 1$- and $Z+\ge2$-jet
events, as measured by CDF at the
Tevatron~\protect\cite{Aaltonen:2007cp}\,, compared against the
next-to-leading-order theoretical calculation from the monte carlo 
MCFM~\protect\cite{Campbell:2002tg}\,.} 
\label{fig:CDFZjet}
\end{center}
\end{figure}
\begin{figure}
\begin{center}
\includegraphics[width=0.99\textwidth]{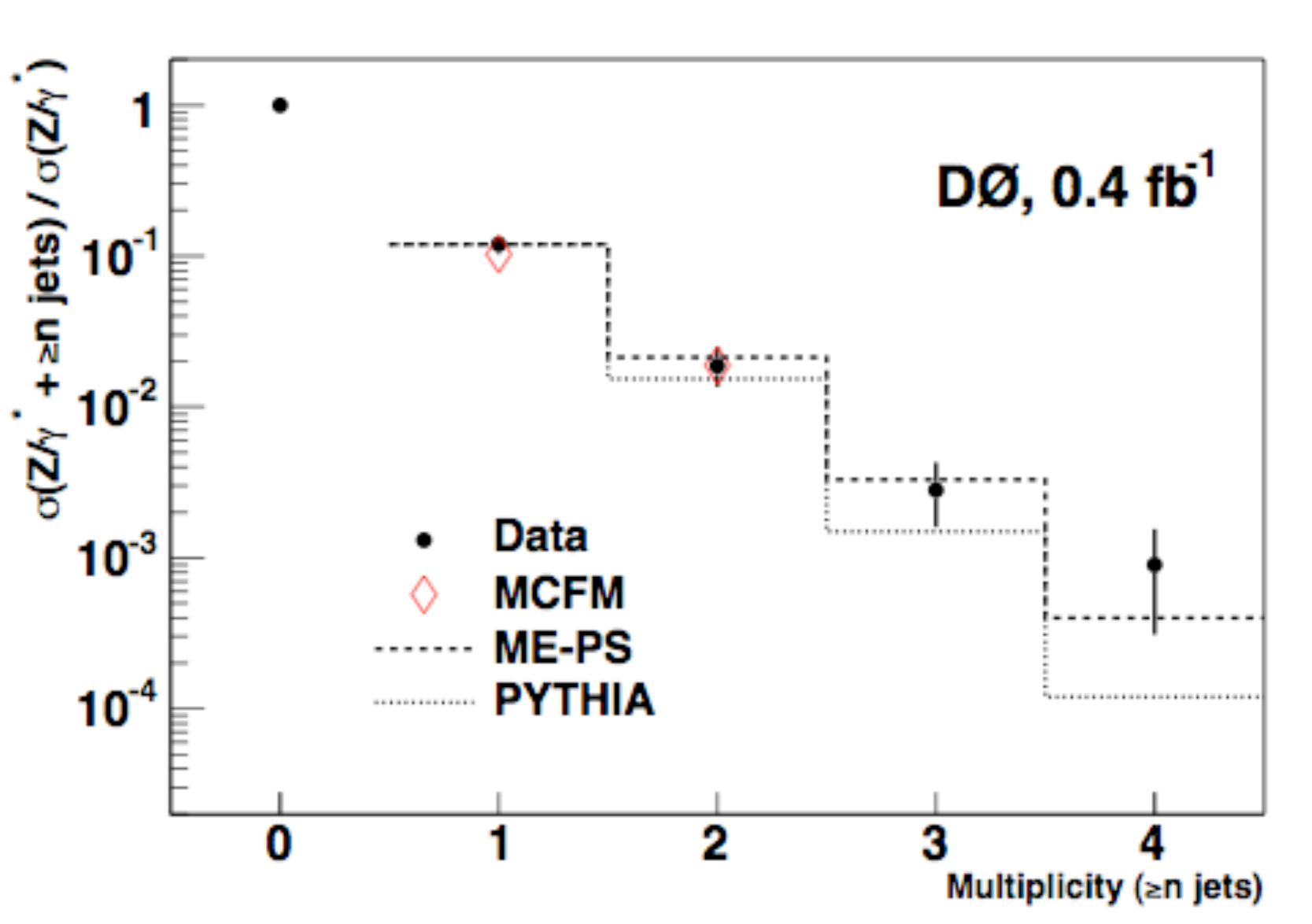}
\caption{Cross section for the production of $Z+$ multijet final
  states, as measured by D0~\protect\cite{Abazov:2006gs}\,, 
compared against the
next-to-leading-order~\protect\cite{Campbell:2002tg} and the leading-order
  matrix-element plus shower 
calculations~\protect\cite{Mrenna:2003if}.}
\label{fig:D0Zjet}
\end{center}
\end{figure}

Once the LHC data will be available, the measurement of the
$(Z\to\ell^+\ell^-)$+jets rates can be used to validate the
extrapolation of the theoretical descriptions from the energy of the
Tevatron to that of the LHC. Few hundred \ipb\ of integrated
luminosity will be enough to determine the overall normalization in
the region of \meff\ dominated by the background (below
1~TeV). Assuming that the shape of the theoretical prediction is
reliable, this will be enough to extend the background estimate at the
higher values of \meff.

\section{$W$+jets}
The processes $W$+jets and $Z$+jets are very similar from the point of
view of QCD. There are minor differences related to the possibly
different initial-state flavour compositions, but the main theoretical
systematics, coming from the renormalization-scale sensitivity due to
the lack of higher-order perturbative corrections, are strongly
correlated. It is therefore reasonable to assume that the clean
measurement of $Z$+jets, with $Z\to \ell^+\ell^-$, should be
sufficient to validate also the calculations of the $W$+jet rates. It
is nevertheless useful to understand the size and features of this
contribution. This is shown in fig.~\ref{fig:w4j} for the
$[W\to\ell\nu]+4$-jet final states, and in fig.~\ref{fig:w3jtau} for the
$[W\to\tau\to\mbox{hadrons}]+3$-jet final states. 
\begin{figure}
\begin{center}
\includegraphics[width=0.99\textwidth]{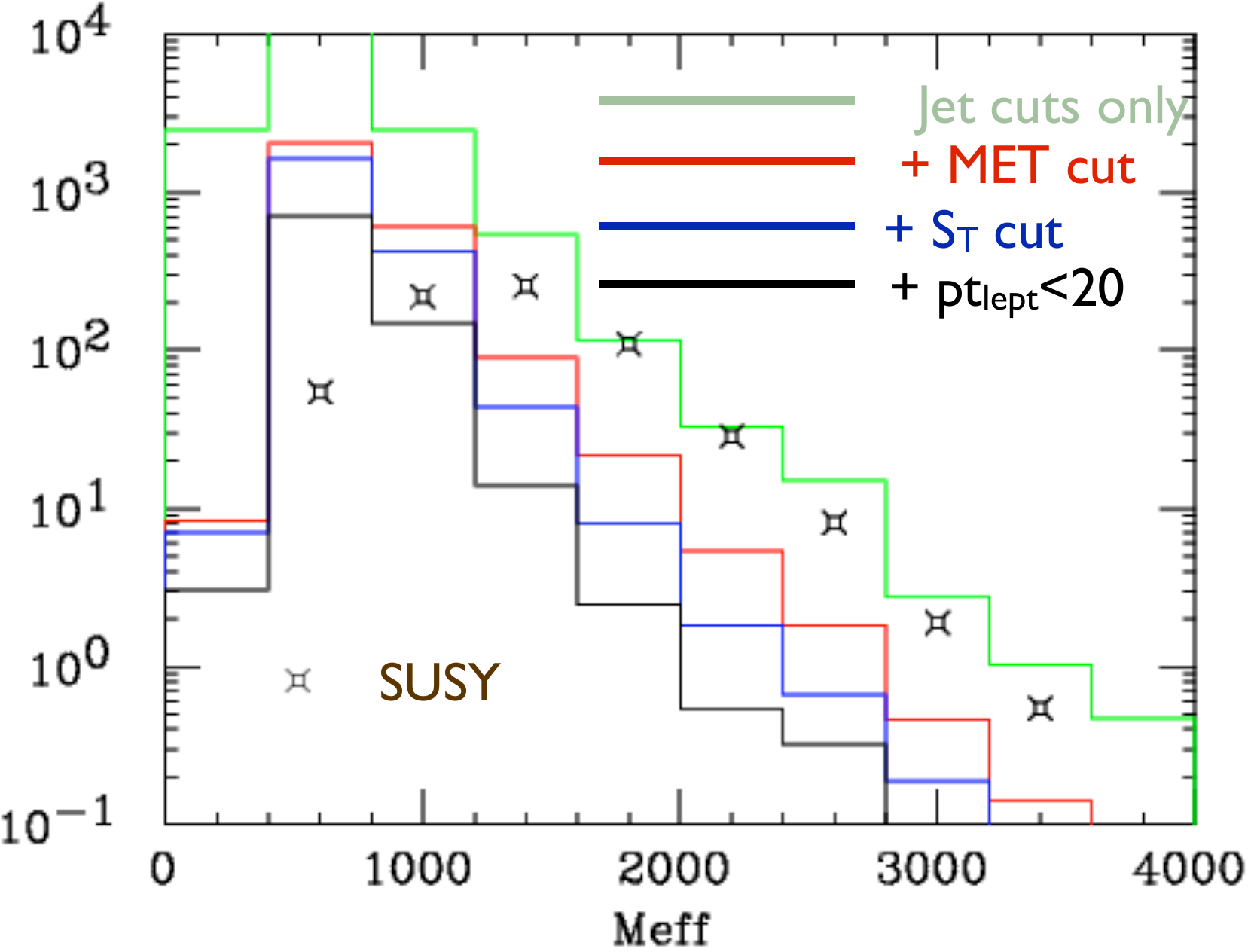}
\caption{Distribution of the effective mass variable, \meff, for the $W$+4~jet
  process, with $W\to\ell\nu$ ($\ell=e,\mu$). 
 The different histograms represent
  the evolution of the background when additional signal cuts are
  imposed to the final state.}
\label{fig:w4j}
\end{center}
\end{figure}

\begin{figure}
\begin{center}
\includegraphics[width=0.99\textwidth]{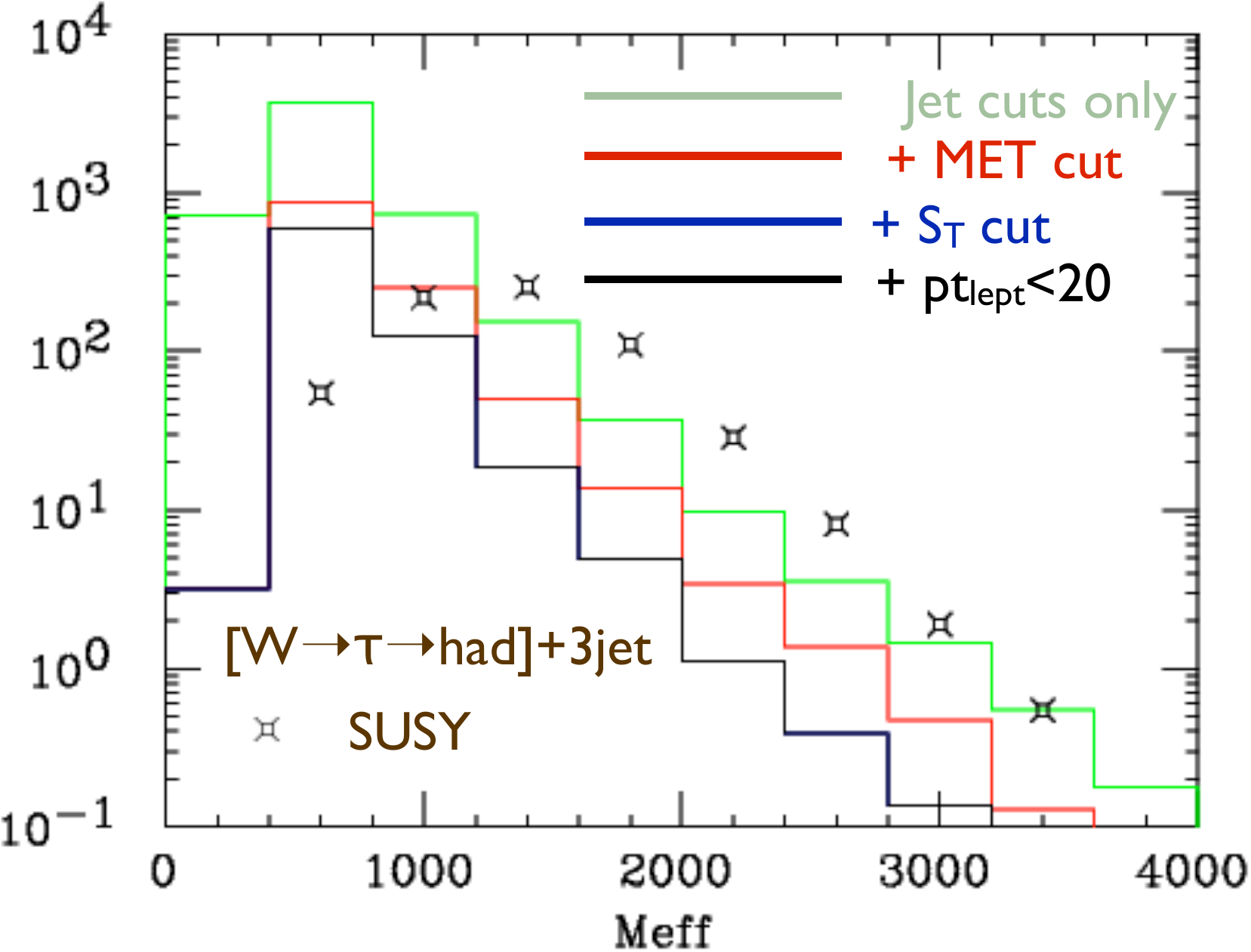}
\caption{Distribution of the effective mass variable, \meff, for the $W$+3~jet
  process, with $W\to\tau\to$ hadrons. 
 The different histograms represent
  the evolution of the background when additional signal cuts are
  imposed to the final state.}
\label{fig:w3jtau}
\end{center}
\end{figure}

The validation of the theoretical predictions for the Tevatron has
been established by a recent measurement at
CDF~\cite{Aaltonen:2007ip}. 
The ratio of the
measured and predicted $W+N$-jet cross
sections, for jets with $\et>25$~\gev, is shown in Fig.~\ref{fig:wjet}.
\begin{figure}
\begin{center}
\includegraphics[width=0.9\textwidth]{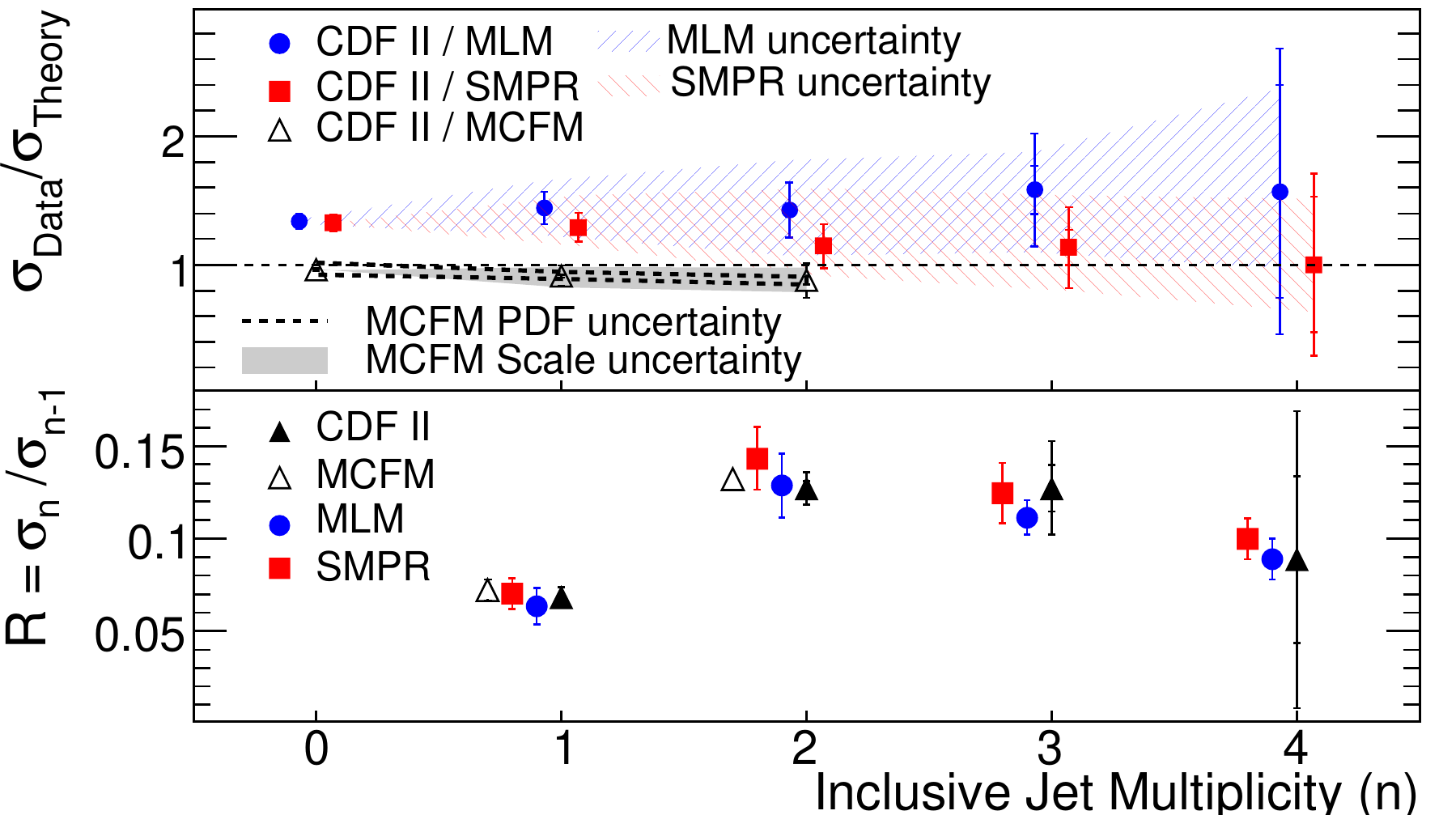}
\caption{Comparison between CDF data and theory for $W+N$ jets cross
  sections~\protect\cite{Aaltonen:2007ip}\,.}
\label{fig:wjet}
\end{center}
\end{figure}
The theoretical predictions include the LO results from
Ref.~\cite{Mangano:2002ea} (labeled as MLM), and from
Ref.~\cite{Mrenna:2003if} (labeled as SMPR), while MCFM refers to the
NLO predictions for the 1- and 2-jet rates from
ref.~\cite{Campbell:2002tg}\,.  The systematic uncertainties of the
individual calculations, mostly due to the choice of renormalization
scale, are shown.  The LO results, which have an absolute
normalization for all $N$-jet values, are in good agreement
with the data, up to an overall $K$ factor, of order 1.4. The
prediction for the ratios of the $N$-jet and $(N-1)$-jet rates is also
in good agreement with the data. The NLO calculations embody the $K$
factor, and exactly reproduce the 1- and 2-jet rates.

Once again, the LHC data will allow a direct check of the correctness
of the theoretical estimates. It should be stressed, however, that the
measurement of the $W$+jets cross section, required to validate the
theoretical calculations, will not be easy. The Tevatron experience
has shown, in fact, that important backgrounds contaminate the $W$+jet
samples. For example, large contributions~\cite{Aaltonen:2007ip} come
from the so-called non-$W$ backgrounds, where the lepton and missing
transverse energy do not originate from a $W$ decay. Possible sources
of such processes include heavy quark (charm and bottom) production,
where the lepton originates from an isolated semileptonic decay of the
heavy quark, and the missing energy from jet mismeasurements and from
the decay to a hard neutrino of the heavy antiquark. Furthermore, for
large jet multiplicity, a large fraction of the $W$+jet signal comes
from $t\bar{t}$ decays, an effect that will be even more pronounced at
the LHC (perversely enough, it can be argued~\cite{Baer:2008dy}
 that supersymmetry itself
could contaminate the measurement of the SM background processes!). 
Therefore, the background to supersymmetry signals has itself
its own backgrounds, and its accurate determination will provide a
challenge by itself.  It is difficult to anticipate the scale of the
challenge, only the direct contact with the data will tell!

\subsection{Theoretical status of $W$+multijet final states}
Due to the importance of $W$+multijet processes as a background to
many analyses, a significant effort has been invested recently in its
theoretical understanding. As mentioned above, NLO calculations are
available for the 1- and 2-jet final
states~\cite{Campbell:2002tg}. These calculations are not available in
the form of full MC event generators, describing the complete
structure of the final states, but provide a reference benchmark for
the results of calculations based on LO matrix elements, merged with
the full shower evolution. Several
groups~\cite{Krauss:2004bs,Krauss:2005nu,Lavesson:2005xu,Maltoni:2002qb, 
Papadopoulos:2005ky,Mangano:2002ea}  have developed tools to
extend the LO predictions to high jet multiplicity, addressing the
problem of merging the higher order processes with the shower
evolution without double counting. The double counting problem refers
to the multiple covering of same phase-space configurations when a jet
can be generated both by the direct matrix element calculation of a
$N$-jet configuration and by the possible hard radiation of a jet
during the shower evolution of a $(N-1)$-parton final state.  
Thorough comparisons have been performed~\cite{Alwall:2007fs} between
the predictions of these tools, resulting in a reliable framework for
the estimates of systematic uncertainties. 
The results of the matrix element evaluation
for these complex processes are all in excellent agreement;
differences in the predictions at the level of hadrons may
arise from the use of different parton-shower approaches, and of
different ways of sharing between matrix elements and shower the task
of describing the radiation of hard jets ({\it merging algorithms}). 
An example of the spread in
the different predictions is shown in Fig.~\ref{fig:pt-lhc}, which
shows the \et\ spectra of the four highest-\et\ jets in $W$+multijet
events at the 
LHC. With the exception of the predictions from one of the codes, all
results are within $\pm 50\%$ of each other, an accuracy sufficient by
itself to establish possible deviations such as those in
Fig.~\ref{fig:susy}.
\begin{figure}
\begin{center}
\includegraphics[width=0.9\textwidth]{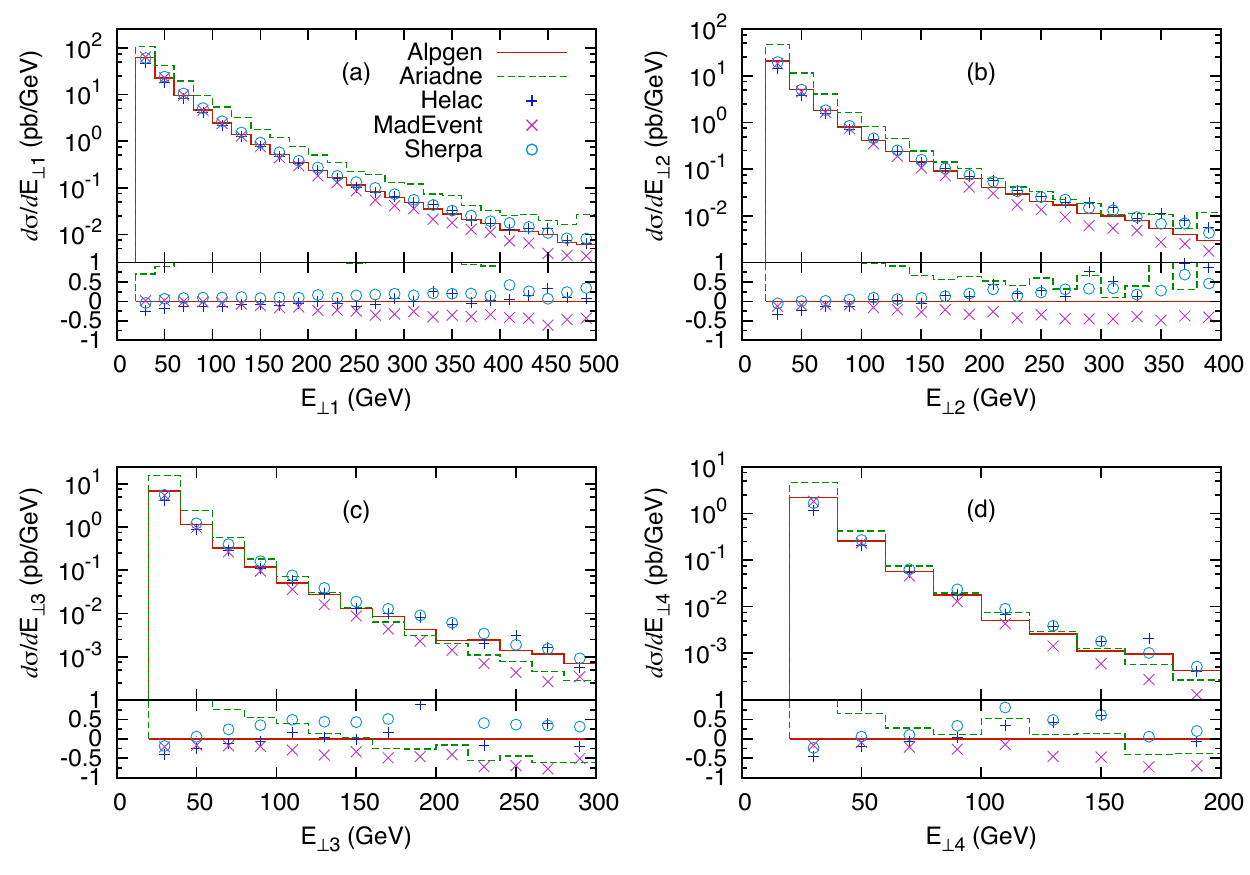}
\caption{Predicted jet \et\ spectra in $W$+jet(s) final states at the
  LHC~\protect\cite{Alwall:2007fs}\,.} 
\label{fig:pt-lhc}
\end{center}
\end{figure}

These differences are of size compatible with the intrinsic
uncertainties of the calculations, given for example by the size of
the bands in Fig.~\ref{fig:wjet}. An alternative picture of the
intrinsic systematics of the ALPGEN predictions for observables like
the \pt\ spectrum of the $W$ boson, the leading-jet pseudorapidity
distribution, and the angular correlations between the two leading
jets, is given in fig.~\ref{fig:alp-ptw-lhc}. Similar results are
obtained for the other codes~\cite{Alwall:2007fs}.  
It is expected that these systematics can be
reduced by tuning the input parameters, like the choice of
renormalization scale, by fitting the data. An accurate determination
of the normalization and shape of the SM background to a
supersymmetric signal could therefore be obtained by analyzing data
control samples. 
\begin{figure}
\begin{center}
\includegraphics[width=0.9\textwidth]{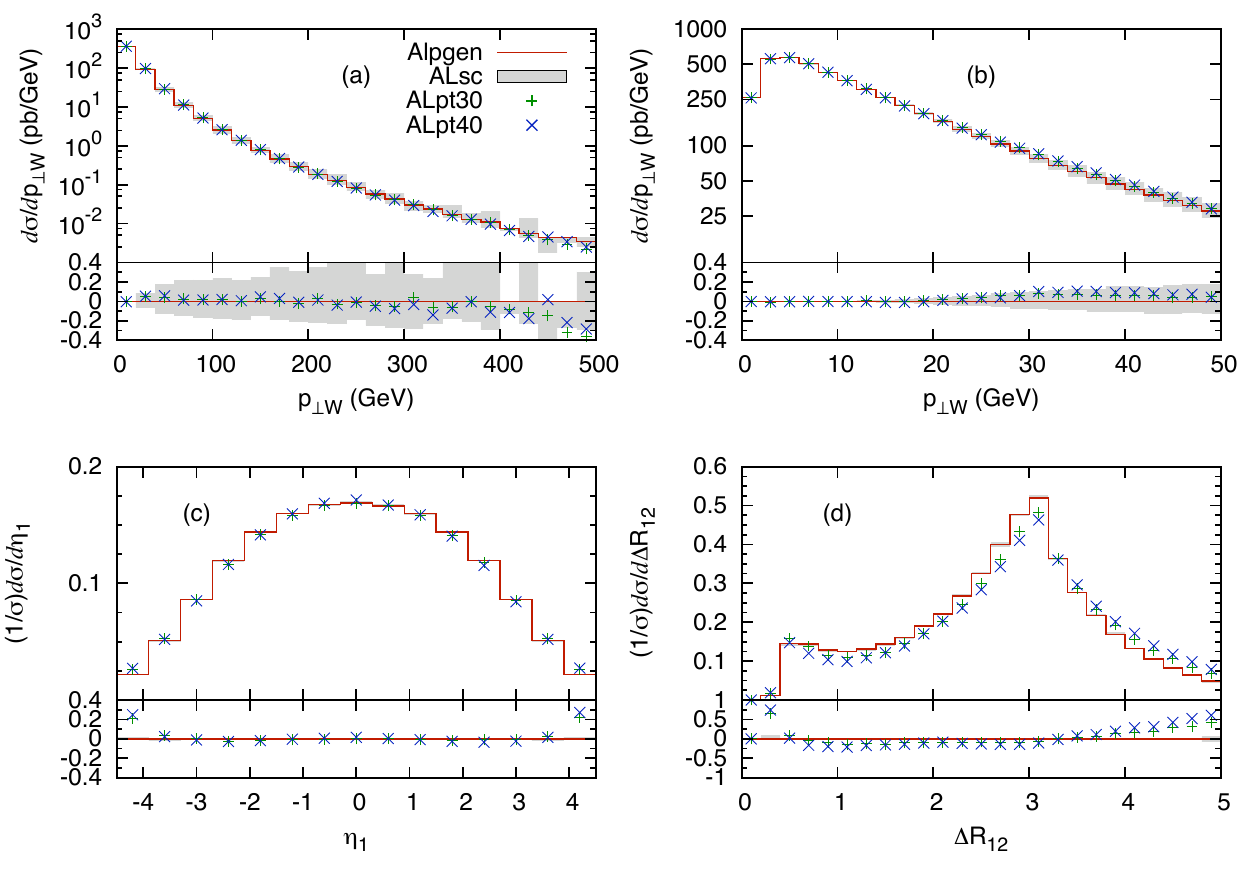}
\caption{Systematic uncertainty for a few $W$+multijet observables,
  estimated from the ALPGEN calculations~\protect\cite{Alwall:2007fs}\,.} 
\label{fig:alp-ptw-lhc}
\end{center}
\end{figure}
A clear path is therefore available to establish the accuracy of the
theoretical tools, and to provide robust background estimates for
searches of anomalies in the multijet plus \met\ final states. 

\section{$t\bar{t}$ production}
With a total cross section of order 1~nb, and a rich set of final states
including leptons, missing energy, jets and heavy
quarks~\cite{Abe:1994st, Abe:1995hr,Abachi:1995iq},
the production of top quarks at the LHC~\cite{Beneke:2000hk}
is one of the potentially largest backgrounds
to almost any type of new phenomena. In the case of the multijet+\met\
signature of supersymmetry, one requests one of the two $W$ bosons from
the $t\bar{t}$ to decay leptonically, and the other hadronically,
leading to a characteristic 4-jet plus \met\ final state.
Early studies of $t\bar{t}$ final
states, nevertheless, suggested that this contribution would be very
small. 
\begin{figure}
\begin{center}
\includegraphics[width=0.99\textwidth]{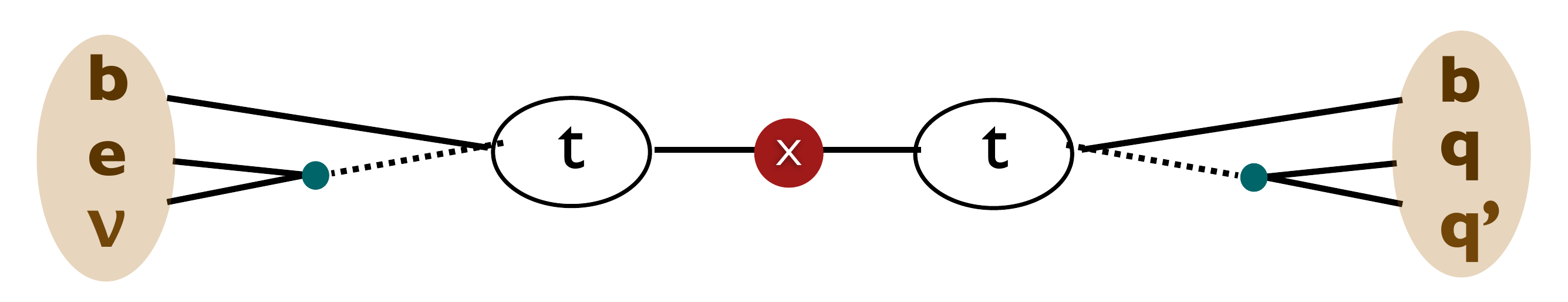}
\includegraphics[width=0.99\textwidth]{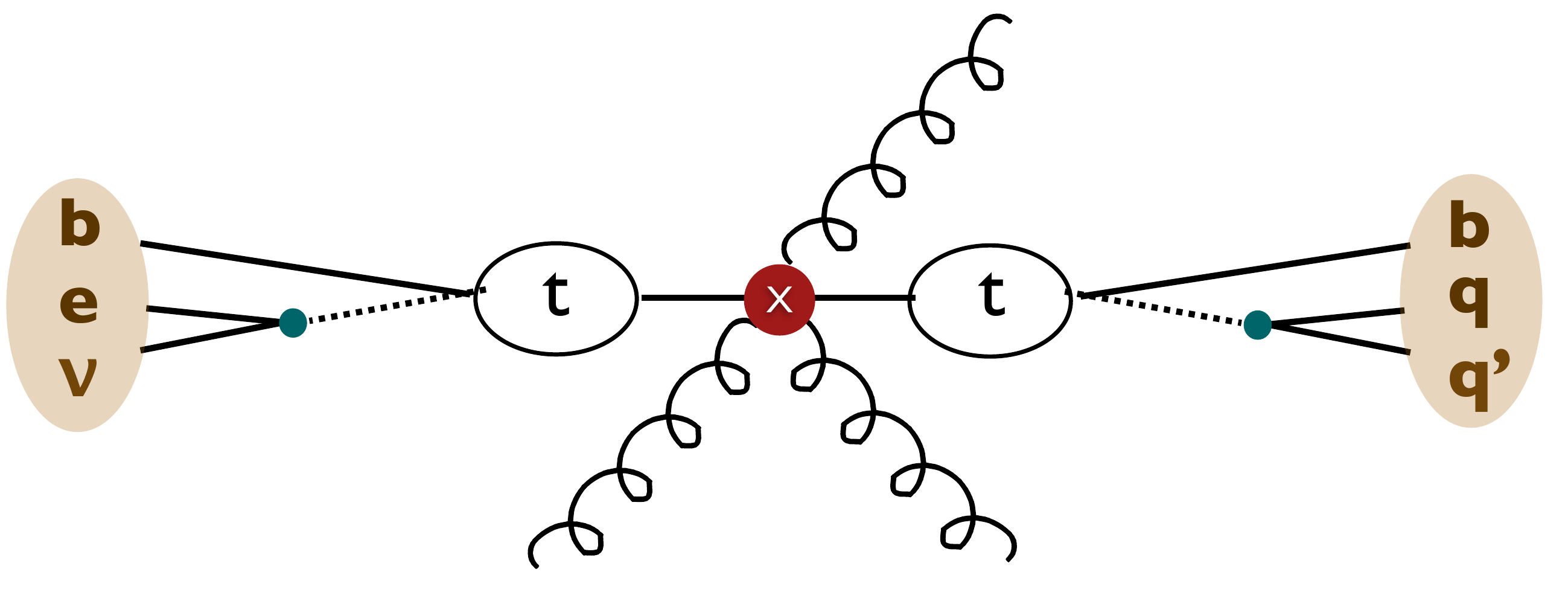}
\caption{$t\bar{t}$ final state at large $p_T$.}
\label{fig:ttbardec}
\end{center}
\end{figure}
Considering only the lowest-order process $pp\to
t\bar{t}$, it is easy to understand
why it should be so. The request of a large value of \meff, in
fact, forces 
the production kinematics into the region where the $t$ and  $\bar{t}$
recoil against each other at very large \pt. This kinematics
leads (see upper panel in fig.~\ref{fig:ttbardec})
 to several consequences: on one side the request for a large
\met\ forces the $W$ to be highly boosted, thus typically leading to a
high-\pt\ lepton as well. On the other, the ``3-jets'' coming from the
hadronic top decay would coalesce into a single fat one, and the
request of having 4 separate jets would not be met. 
This implies that, in order to satisfy the supersymmetry-selection
cuts, one needs additional hard jets not coming from the top decays
(see lower panel of fig.~\ref{fig:ttbardec}). 
The dominance of these higher-order processes is confirmed by the
calculation, as shown in fig.~\ref{fig:tt}, where we present the \meff\ spectra
of events with 1, 2 and 3 hard partons in addition to the $t\bar{t}$
decays products. Notice that these
high-jet-multiplicity final states would typically be underestimated by the
standard parton-shower approximation. As in the case of the
$Z/W+$multijet final states, a reliable background estimate requires
therefore the use of higher-order matrix element codes. 

\begin{figure}
\begin{center}
\includegraphics[width=0.99\textwidth]{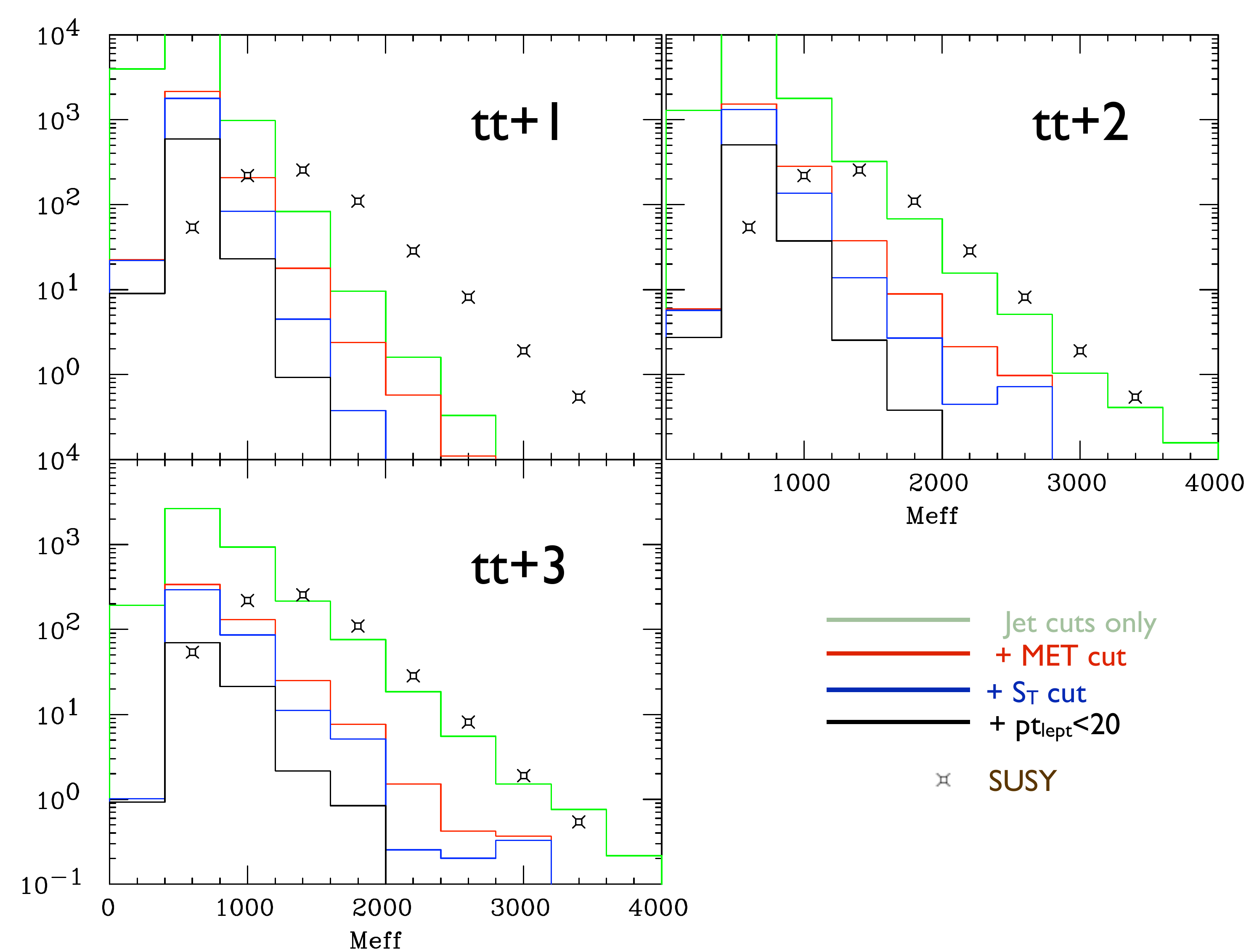}
\caption{Contributions to the \meff\ distribution coming from
  production of $t\bar{t}$ pairs in association with 1, 2 and 3 extra
  hard partons.}
\label{fig:tt}
\end{center}
\end{figure}

\subsection{Theoretical status of $t\bar{t}$+multijet final states}
The theoretical prediction for 
inclusive $t\bar{t}$ production have already been well tested at the
Tevatron~\cite{Sorin:2007zz,Pleier:2008hp,Yao:2006px}\,. 
For example, NLO calculations~\cite{Nason:1987xz,Beenakker:1988bq}, 
enhanced
by the resummation of leading~\cite{Catani:1996yz} 
and subleading Sudakov
logarithms~\cite{Bonciani:1998vc,Cacciari:2003fi,Cacciari:2008zb} 
or by the inclusion of classes of NNLO
 terms~\cite{Kidonakis:2003qe,Moch:2008qy,Kidonakis:2008mu} 
predict the total cross
section with an accuracy consistent with the experimental
uncertainty~\cite{:2007qf,Abulencia:2006kv,Acosta:2005am,Acosta:2004uw,:2007bu,
Abazov:2007kg, Abazov:2006yb,Abazov:2006ka,Abazov:2005ey},
as shown by fig.~\ref{fig:ttsig}.
\begin{figure}
\begin{center}
\includegraphics[width=0.99\textwidth]{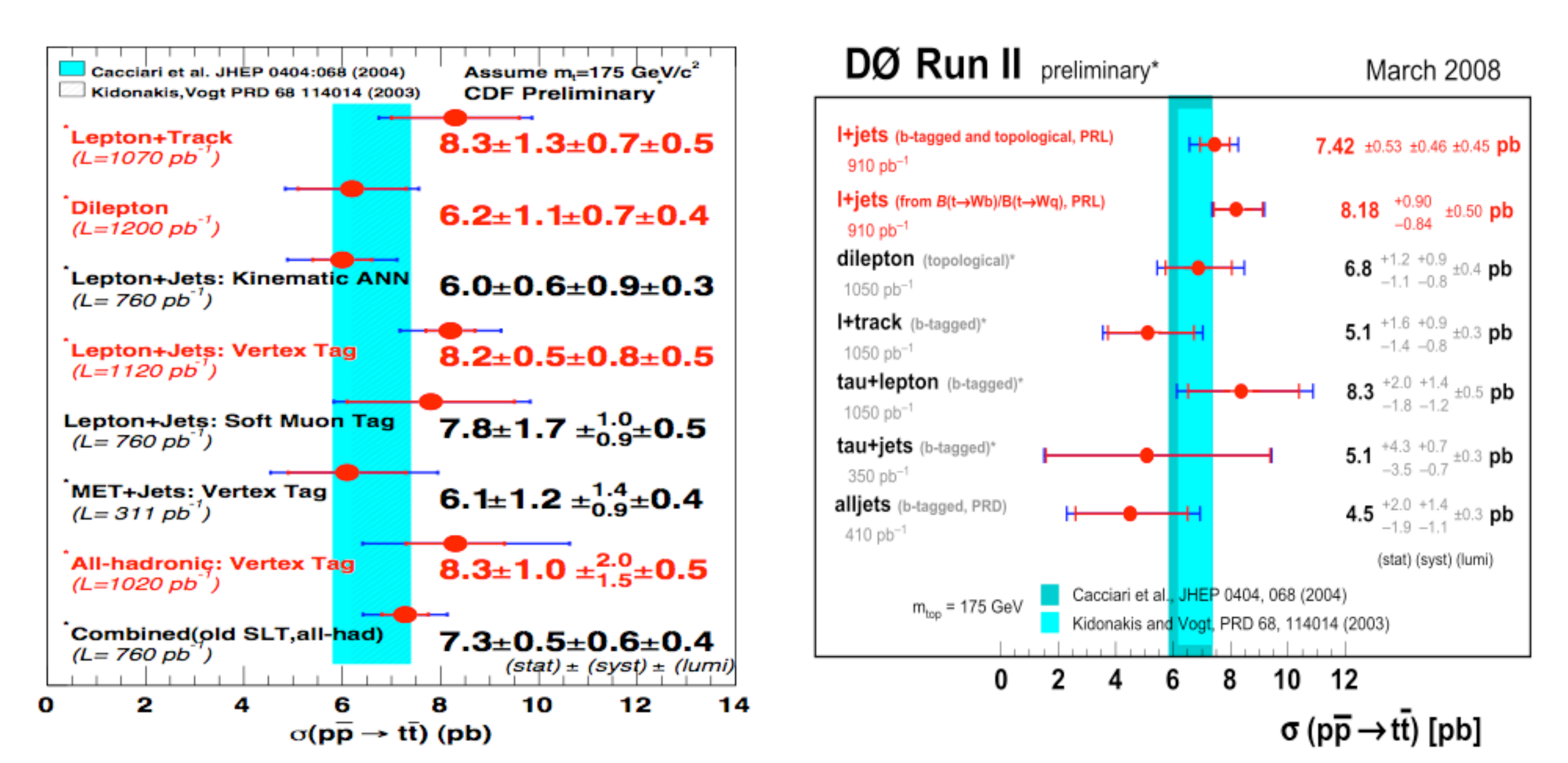}
\caption{$t\bar{t}$ cross section measurements at the Tevatron from
  CDF (left) and D0 (right). Summary taken from~\protect\cite{Pleier:2008hp}.}
\label{fig:ttsig}
\end{center}
\end{figure}
The predictions for the LHC are expected to be equally
accurate~\cite{Cacciari:2008zb,Moch:2008qy,Kidonakis:2008mu}, if not
more, since the main source of uncertainty, the parton distribution
functions, fall at the LHC in a range of $x$ values where they are
known with precision better than at the
Tevatron~\cite{Pumplin:2002vw,Martin:2001es}.
Recent progress towards a complete NNLO
calculation~\cite{Czakon:2007ej,Czakon:2007wk,Czakon:2008zk,Korner:2005rg,
  Korner:2008bn,Dittmaier:2008jg} 
will likely push the ultimate accuracy of the total cross section to
the level of few percent.

The extreme kinematics of $t\bar{t}$ final states responsible for the
background to supersymmetry searches, however, cannot be tested at the
Tevatron, due to the limited statistics and phase-space. It is only
with the LHC data that compelling tests will become possible. Except
for $t\bar{t}+1$ jet final states~\cite{Dittmaier:2008jg}, predictions
for the associated production of $t\bar{t}$ and several jets are only
available using LO matrix element calculations, merged with
parton shower evolution codes. Where possible, these LO tools have
been validated against the existing NLO+shower descriptions. The NLO
corrections to $t\bar{t}$ production have been
incorporated~\cite{Frixione:2003ei,Frixione:2007nw}  in fact in
the HERWIG~\cite{Corcella:2000bw} and PYTHIA~\cite{Sjostrand:2003wg}
 Monte Carlos, allowing for a complete description, at NLO accuracy,
 of the inclusive $t\bar{t}$ final states. Predicted properties 
such as the transverse momentum distribution or the invariant
mass of the $t\bar{t}$ pair, agree very well with those extracted from
the LO calculations merged with parton
showers~\cite{Mangano:2006rw}\,, as shown in
fig.~\ref{fig:ttdist}. Differences between the two calculations
emerge when considering the production rate of several additional
jets, but remain relatively modest, as shown in
fig.~\ref{fig:ttjet}. Specific studies dealing with the high-\pt\
regions characteristic of the supersymmetry searches, however, have
not been carried out as yet. Once again, only the direct comparisons with the
LHC data will provide the definitive validation of these calculations.
\begin{figure}
\begin{center}
\includegraphics[width=0.49\textwidth]{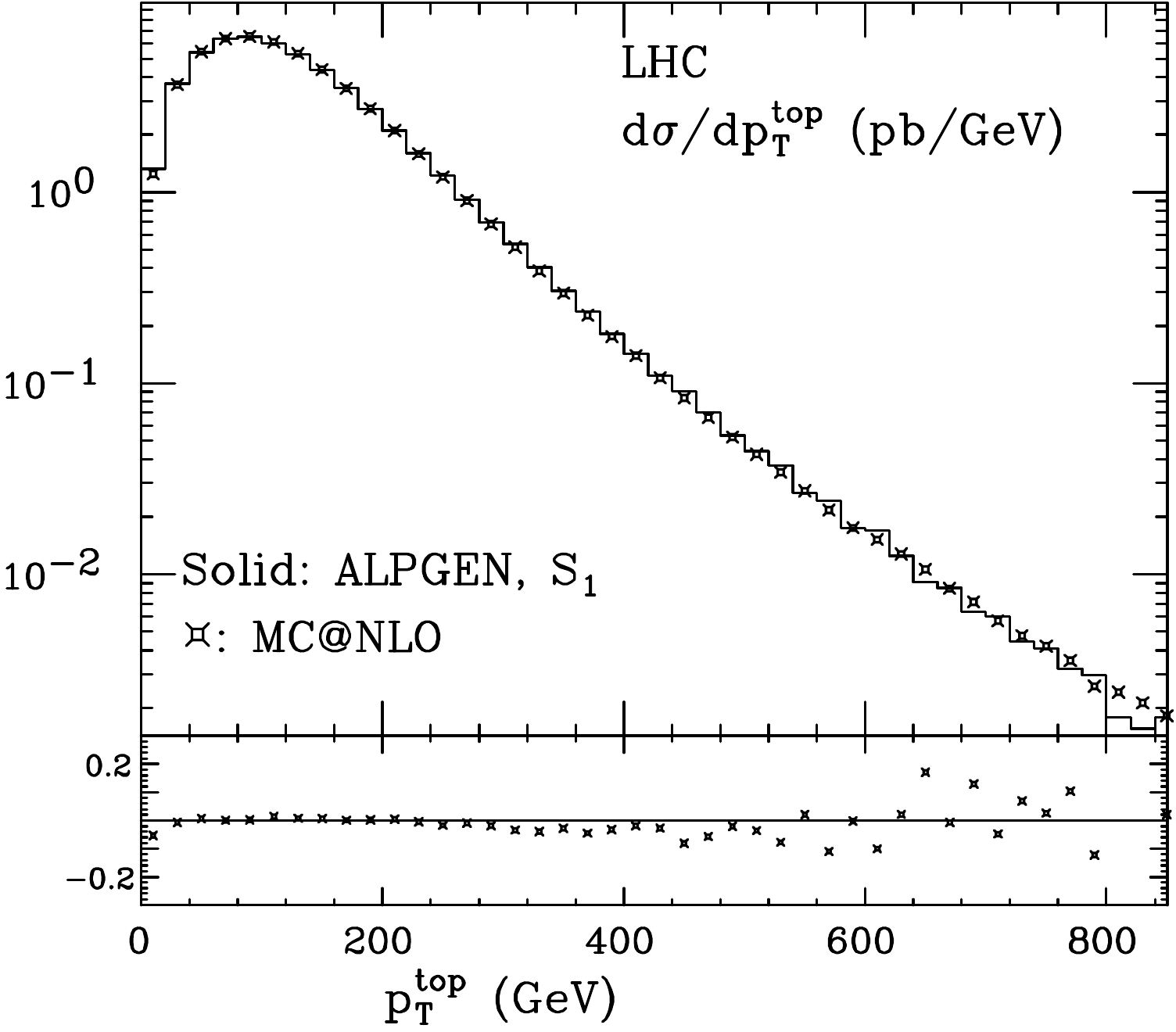}
\hfill
\includegraphics[width=0.49\textwidth]{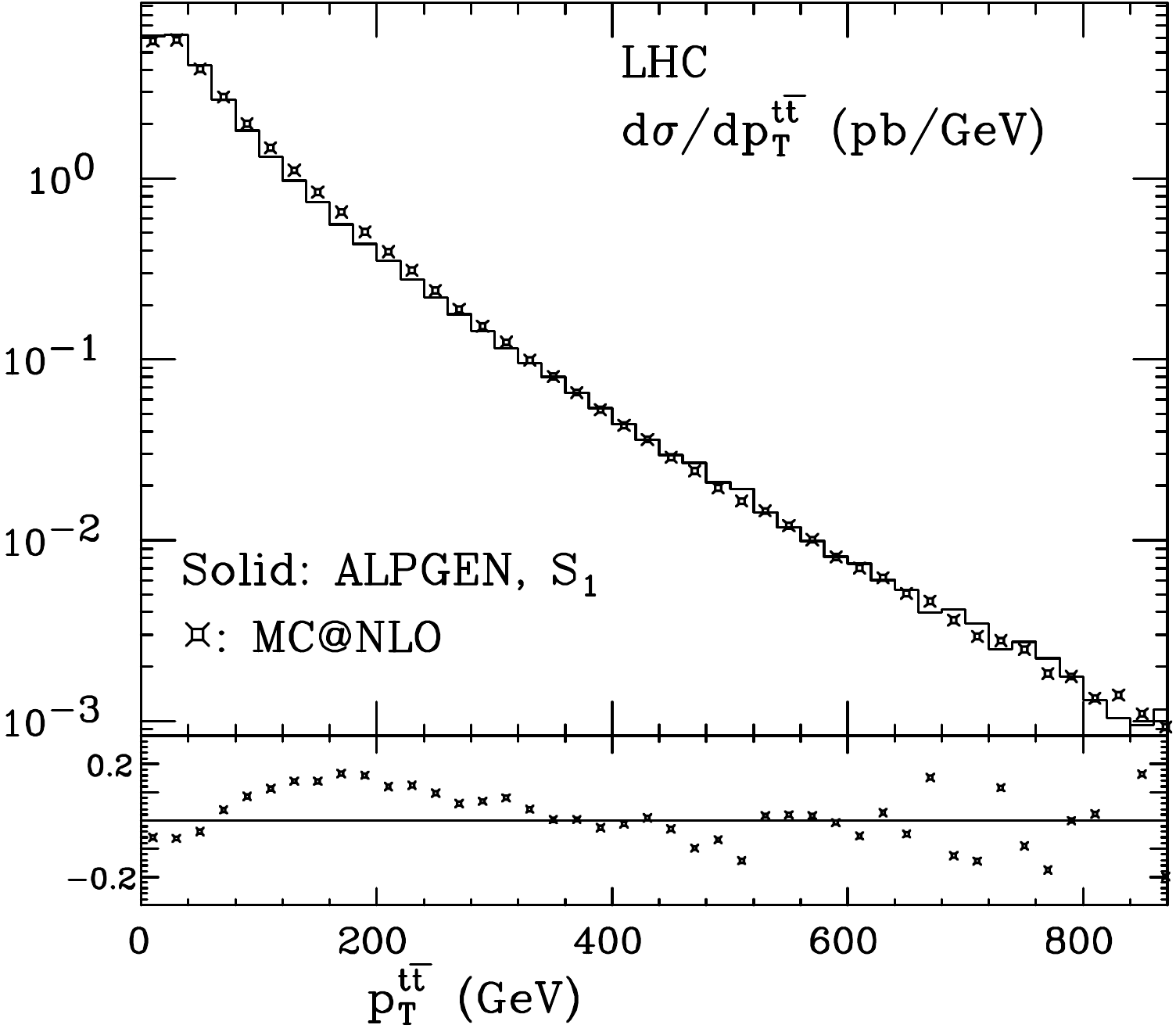}
\caption{Inclusive spectrum of the top quark (left), and \pt\ of the
  $t\bar{t}$ pair (right) as described by the
  MC@NLO (points) and by ALPGEN LO+HERWIG calculations.}
\label{fig:ttdist}
\end{center}
\end{figure}
\begin{figure}
\begin{center}
\includegraphics[width=0.9\textwidth]{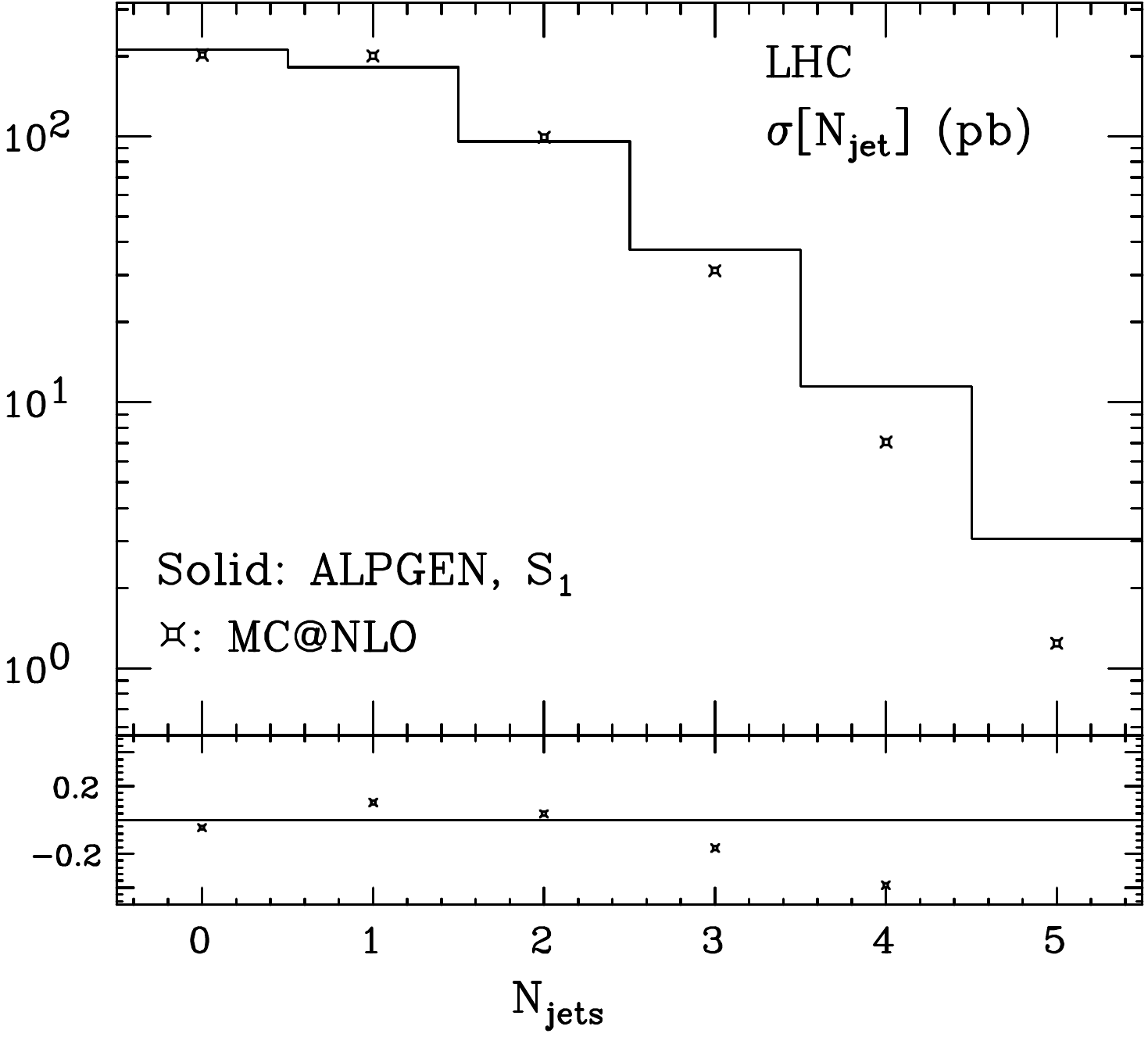}
\caption{Multiplicity of jets associated to 
  $t\bar{t}$ pairs, as described by the
  MC@NLO (points) and by ALPGEN leading-order+HERWIG calculations.}
\label{fig:ttjet}
\end{center}
\end{figure}

\section{QCD multijets}
The large energy available in LHC collisions can
allow for the loss of significant amounts of transverse energy via
jets emitted at very large rapidity, as shown in fig.~\ref{fig:jetmet}. 
\begin{figure}
\begin{center}
\includegraphics[width=0.9\textwidth]{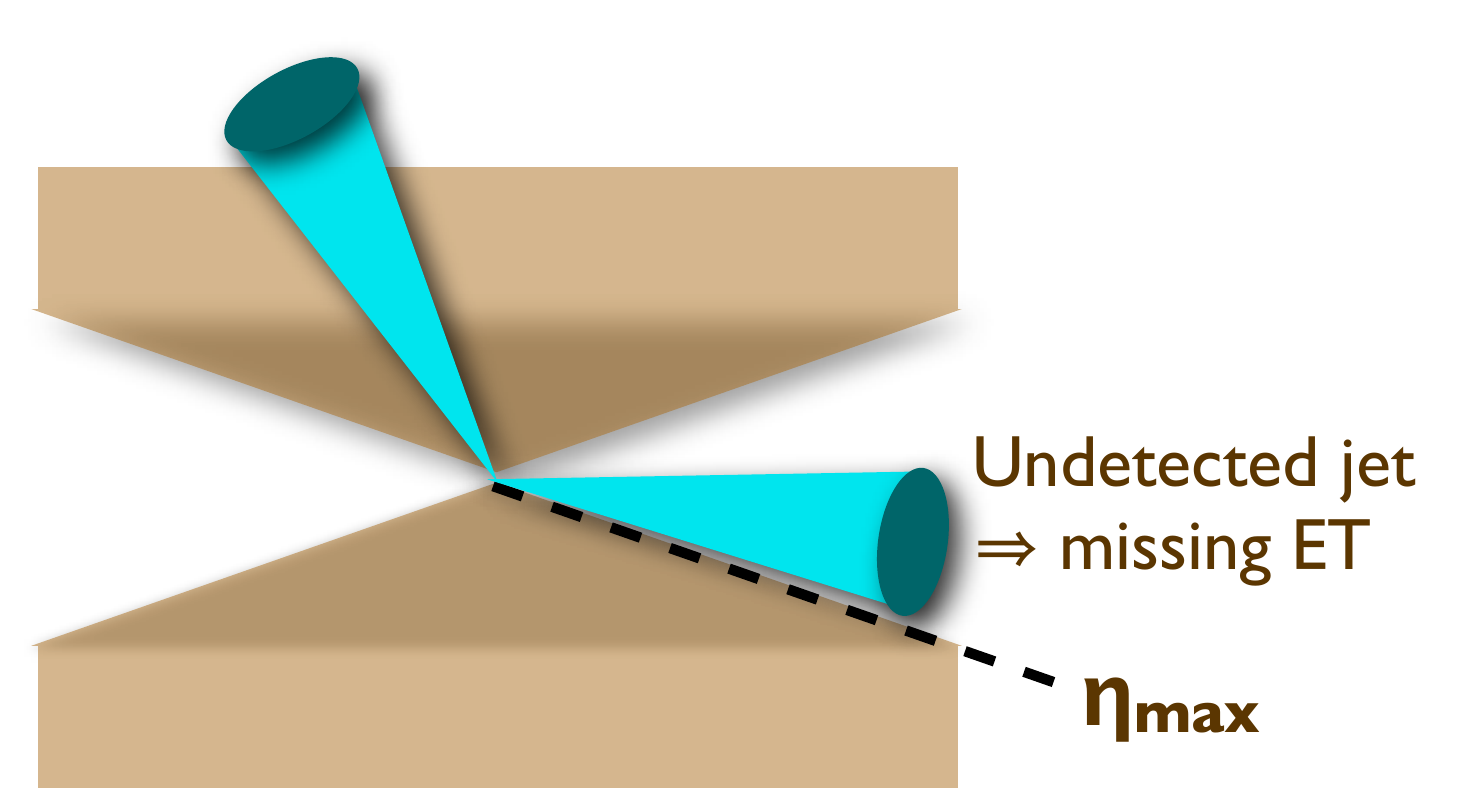}
\caption{Limited calorimetric coverage as a source of missing
  transverse momentum in multijet final states.}
\label{fig:jetmet}
\end{center}
\end{figure}
A quantitative estimate of this effect is shown in
fig.~\ref{fig:dijetmet}. We show here the ratio
\be
\frac{\sigma_{jj}(\met>E_{T0}; ~ \vert\eta\vert>\eta_{max})}{\sigma_{pp}}
\ee
where $\sigma_{pp}$ is the total $pp$ cross section, and the numerator
is the cross section to produce a dijet pair, with one of the two jets
escaping detection due to its large pseudorapidity
$\vert\eta\vert>\eta_{max}$. These final 
states would appear as a 1-jet event, with missing \et. If such events
were to overlap with a 3-jet event, they would lead to a 4-jet+\met\
signature, faking the supersymmetry signal. At high luminosity, the
large number of additional $pp$ interactions would amplify the
probability that one such jet+\met\ event overlapped with whatever
primary multijet final state. The results of fig.~\ref{fig:jetmet}
show that a calorimetric coverage out to $\eta$ of
order 3 of 4 would  lead to a large \met\ signal. At the
highest expected luminosities, $L\sim 10^{34}$cm$^{-2}$s$^{-1}$, the number of
overlapping events is of the order of 20, and the probability of
$\met\gsim 100$~GeV can become of O$(10^{-4})$, thus leading to a background
as large as the signal. For this reason the LHC calorimeters extend
out to about $\eta=5$. The current theoretical calculations of
multijet final states are based on LO matrix elements, merged with
shower MC. In the case of ALPGEN, this extends out to multiplicities
of about 6, thus suitable for the study of supersymmetry
backgrounds. These calculations were tested during the run~1 of the
Tevatron, showing a good agreement~\cite{Abe:1996nn}. The
early LHC data taking, with a low luminosity and a small number of
overlapping $pp$ collisions, will provide a robust validation of the
calculations. Other instrumental sources of \met, as e.g. the
non-gaussian tails of the jet-energy resolution, will be monitored
studying the energy balance of $\gamma$-jet and $Z$-jet events, as
discussed in the case of the Tevatron analyses in ref.~\cite{Bhatti:2005ai}.

\begin{figure}
\begin{center}
\includegraphics[width=0.9\textwidth]{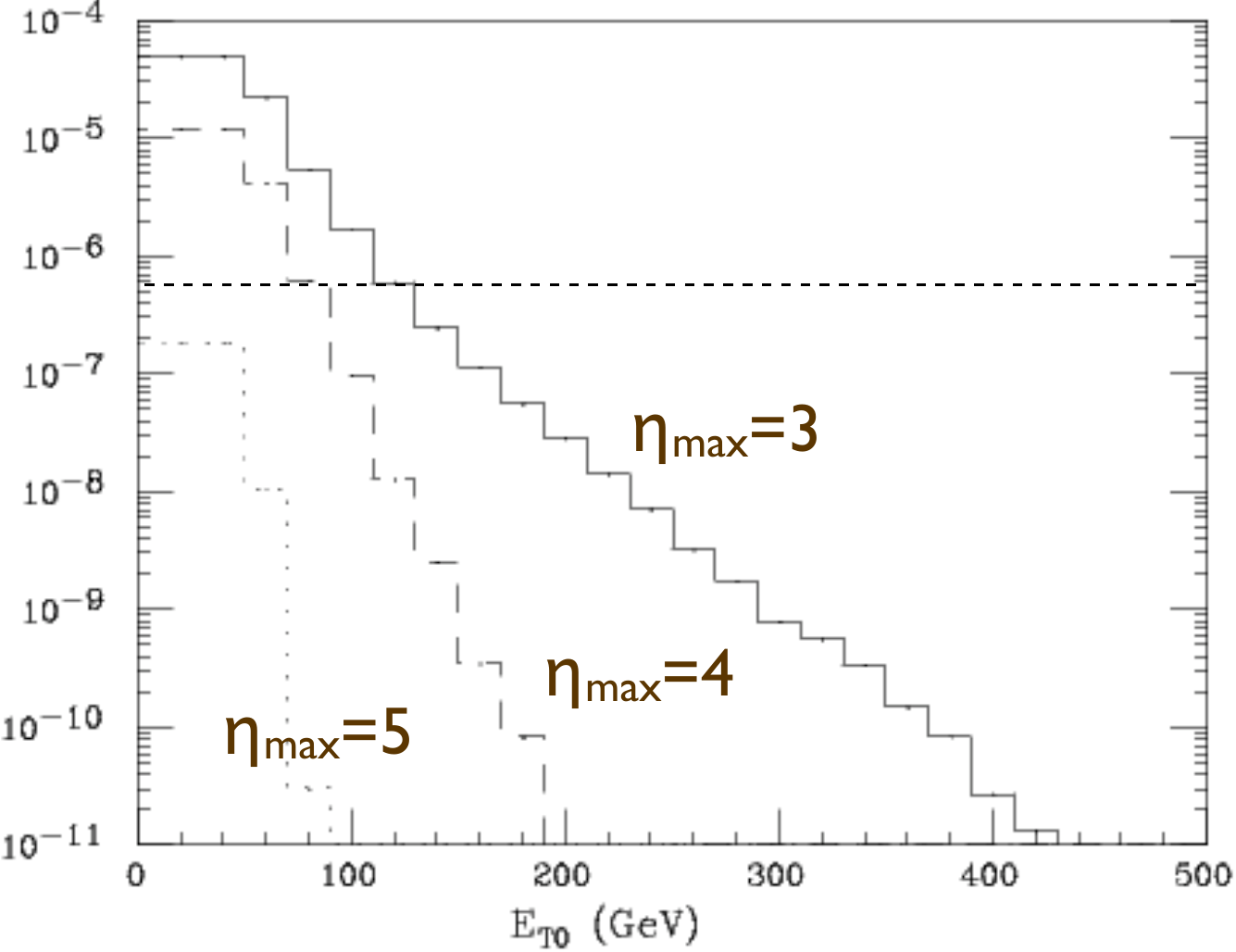}
\caption{Fraction of the total $pp$ cross section 
 with missing transverse energy $\met>E_{T0}$, due to the loss
  of a jet at pseudorapidity
 $\eta>\eta_{max}$, for various values of $\eta_{max}$. 
The horizontal
dashed line corresponds to the rate of inclusive
$W$ events decaying leptonically.}
\label{fig:dijetmet}
\end{center}
\end{figure}

\section{Conclusions}
Advanced MC tools for the description of the SM, and for the isolation
of possible new physics at the LHC, are becoming mature.  Validation
and tuning efforts are underway at the Tevatron, and show that a solid
level of understanding of even the most complex manifestations of the
SM are well under control.  The extrapolation of these tools to the
energy regime of the LHC is expected to be reliable, at least in the
domain of expected discoveries, where the energies of individual
objects (leptons, jets, missing energy) are of order 100~GeV and more. However,
the consequences of interpreting possible discrepancies as new physics
are too important for us to blindly rely on our faith in the goodness of the
available tools.  An extensive and coherent campaign of MC testing,
validation and tuning at the LHC will therefore be required.  Its
precise definition will probably happen only once the data are
available, and the first comparisons will give us an idea of how far
off we are and which areas require closer scrutiny.

Ultimately the burden, and the merit, of a discovery should and will
only rest on the experiments themselves! The data will provide the
theorists guidance for the improvement of the tools, and the analysis
strategies will define the sets of control samples that can be used to
prepare the appropriate and reliable use of the theoretical
predictions.

Aside from the discovery of anticipated objects like the $W$, $Z$ and
the top, we have never faced with high-energy colliders the concrete
situation of a discovery of something beyond the expected. In this
respect, we are approaching what the LHC has in store for us without a
true experience of discovering the yet unknown, and we should
therefore proceed with great caution.  All apparent instances of
deviations from the SM emerged so far in hadronic or leptonic
high-energy collisions have eventually been sorted out, thanks to
intense tests, checks, and reevaluations of the experimental and
theoretical systematics.  This shows that the control mechanisms set
in place by the commonly established practice are very robust.

Occasionally, this conservative approach has delayed in some areas of
particle physics the acceptance of true discoveries, as in the case of
Davies's neutrino mixing, and as might turn out to be the case for the
muon anomaly. But it has never stopped the progress of the field, on
the contrary, it has encouraged new experimental approaches, and has
pushed theoretical physics to further improve its tools.

The interplay between excellent experimental tools, endowed with the
necessary redundancy required to cross-check odd findings between
different experiments and different observables, and a hard-working
theoretical community, closely interacting with the experiments to
improve the modeling of complex phenomena, have provided one of the
best examples in science of responsible and professional {\it modus
operandi}. In spite of all the difficult challenges that the LHC will
pose, there is no doubt in my mind that this articulated framework of
enquiry into the yet unknown mysteries of nature will continue
providing compelling and robust results.

\section*{Acknowledgements}
This work is supported in part by the European Community's 
Marie-Curie Research Training Network HEPTOOLS under contract 
MRTN-CT-2006-035505.

\end{document}